\PassOptionsToPackage{nopatch=footnote}{microtype}
\documentclass[screen,sigplan,10pt]{acmart}

\acmYear{2026}\copyrightyear{2026}
\acmConference[SOSP '26]{Symposium on Operating Systems Principles}{September 29--October 2, 2026}{Prague, Czech Republic}
\acmBooktitle{Symposium on Operating Systems Principles (SOSP '26), September 29--October 2, 2026, Prague, Czech Republic}
\acmDOI{10.1145/3830418.3843902}
\acmISBN{979-8-4007-2585-2/26/09}

\setcopyright{cc}
\setcctype[4.0]{by}
\copyrightyear{2026}
\acmYear{2026}
\usepackage{amsmath,pifont}
\usepackage{xspace, xurl}
\usepackage[table]{xcolor}
\usepackage{gnuplot-lua-tikz}
\usepackage{hyperref,fancyhdr, framed, changepage}
\usepackage{verbatim}
\usepackage[font=small]{caption}
\usepackage{subcaption}
\usepackage[textsize=tiny]{todonotes}
\usepackage{tabularx, newfloat, placeins}
\usepackage{balance}
\usepackage{enumitem}
\usepackage{listings}
\usepackage{musicography}
\usepackage[ruled]{algorithm2e}
\SetAlCapFnt{\small}
\SetAlCapNameFnt{\small}
\usepackage{multirow}
\usepackage{makecell}

\usepackage[most]{tcolorbox}

\usepackage[underline=true]{pgf-umlsd}
\usetikzlibrary{arrows.meta,calc,fit,shapes,decorations.pathreplacing}

\hypersetup{
    colorlinks,
    linkcolor={red!50!black},
    citecolor={blue!50!black},
    urlcolor={blue!80!black}
}

\colorlet{DarkRed}{red!50!black}

\makeatletter
\DeclareFontEncoding{LS1}{}{}
\DeclareFontEncoding{LS2}{}{\noaccents@}
\DeclareFontSubstitution{LS1}{stix}{m}{n}
\DeclareFontSubstitution{LS2}{stix}{m}{n}
\makeatother

\let\mathscr\undefined
\let\mathcal\undefined
\DeclareMathAlphabet\mathscr{LS1}{stixscr}{m}{n}
\SetMathAlphabet\mathscr{bold}{LS1}{stixscr}{b}{n}
\DeclareMathAlphabet\mathcal{LS2}{stixcal}{m}{n}
\SetMathAlphabet\mathcal{bold}{LS2}{stixcal}{b}{n}

\DeclareFontFamily{T1}{zi4n}{}
\DeclareFontShape{T1}{zi4n}{m}{n}{<-> t1-zi4nr-0}{}
\DeclareFontShape{T1}{zi4n}{b}{n}{<-> t1-zi4nb-0}{}
\DeclareFontShape{T1}{zi4n}{bx}{n}{<-> ssub * zi4n/b/n}{}

\newcommand{\xmark}{\ding{53}}
\lstdefinestyle{substrate}{
    language=Python,
    morekeywords={view,make_shared,make_local,matmul,tag,const,Tensor,
      syncthreads,u32,u64,u128,u256,bf16,f32}
}

\newcommand{\inlinesection}[1]{\par\addvspace{0.3\baselineskip plus 0.05\baselineskip minus 0.05\baselineskip}\noindent{\emph{#1}}}

\newcommand{\inlinesectiontighter}[1]{\par\addvspace{0.2\baselineskip plus 0.05\baselineskip minus 0.05\baselineskip}\noindent{\emph{#1}}}

\title{\textsc{Ave}: Guiding Agentic GPU Optimization Using Data-Flow Invariants}
\newcommand{\system}{\textsc{Ave}\xspace}

\newcommand{\authorhspace}{\hspace{2ex}}

\author{
{\rm Haohui Mai\textsuperscript{$\scriptscriptstyle\musEighth,\mkern2mu1$}} \authorhspace
{\rm Xiaoyan Guo\textsuperscript{4}} \authorhspace
{\rm Xiangyun Ding\textsuperscript{$\scriptscriptstyle\musEighth,\mkern2mu5$}}  \authorhspace
{\rm Daifeng Li\textsuperscript{1}} \authorhspace
{\rm Qiuchu Yu\textsuperscript{4}} \\[0.4ex]
{\rm Chenzhun Guo\textsuperscript{4}}  \authorhspace
{\rm Cong Wang\textsuperscript{2}} \authorhspace
{\rm Jiacheng Zhao\textsuperscript{4}} \authorhspace
{\rm Christos Kozyrakis\textsuperscript{3}} \authorhspace
{\rm Binhang Yuan\textsuperscript{1}}
}

\affiliation{
\institution{
{CausalFlow Inc.\textsuperscript{$\scriptscriptstyle\musEighth$}} \authorhspace
{HKUST\textsuperscript{1}} \authorhspace
{Tsinghua University}\textsuperscript{2}  \authorhspace
{Stanford University}\textsuperscript{3} \authorhspace
{UCAS\textsuperscript{4}} \authorhspace
{UC Riverside\textsuperscript{5}} \authorhspace
}
\city{}
\country{}
}

\newcommand{\floatrule}{\hrulefill}
\begin{document}

\begin{abstract}
LLM coding agents can generate correct GPU kernels, yet their performance still
lags behind expert libraries. Achieving peak GPU throughput requires coordinating
low-level transformations such as shared-memory staging, software pipelining,
and instruction scheduling. Unit tests and performance profiles provide only
sparse end-to-end feedback, leaving agents unable to localize violations of the
global constraints that these transformations must preserve.

We present \system, an agentic framework that uses \emph{data-flow invariants}
as compile-time guardrails for kernel optimization. These invariants specify
relationships among data values that must hold throughout execution. \system's
tile-based Pythonic DSL exposes hardware instructions and compiler policies
while abstracting complex memory layout encodings as tiles. Tag functions assign symbolic labels
derived from selected logical coordinates and expressions. The compiler
propagates these labels through data and control flow, and tag assertions enforce
required relationships among them at use sites. The compiler checks these assertions with a tractable,
flow-sensitive, path-insensitive analysis backed by an SMT solver, returning
concrete assignments that witness abstract violations and guide repairs with no
runtime overhead. An in-context reinforcement learning (ICRL) planner proposes
optimizations from a curated knowledge base. A lowering agent implements them
and instantiates their invariants.

We evaluate \system on AMD MI300X across GEMM, flash attention, and MoE, which
together account for up to 90\% of GPU time in LLM inference. With GPT-5.6 Sol,
its optimized kernels achieve 89--99\% of the effective throughput of
state-of-the-art hand-optimized libraries and improve geometric-mean throughput
by $1.62$--$1176\times$ over uncontaminated agentic baselines. On 200 KernelBench tasks,
with GPT-5.6 Sol and in-context DSL examples, \system produces valid kernels
within three attempts for 100\% of Level~1 and 88\% of Level~2 problems. \system
is open source at
\url{https://github.com/causalflow-ai/avelang}.
\end{abstract}

\maketitle
\pagestyle{plain}

% Display page numbers
%\thispagestyle{empty}
%\pagestyle{empty}

%\thispagestyle{fancy}
%\pagestyle{fancy}
%\fancyhead{}
%\fancyfoot{}
%\fancyhf[CH,CF]{CONFIDENTIAL -- DO NOT DISTRIBUTE}
%\fancyhf[RF]{\thepage}

% !Tex Root = ./paper.tex
\section{Introduction}
\label{s:intro}

Modern large language models (LLMs)~\cite{OpenAI:2026a} increasingly exhibit
broad, human-level capabilities and may be approaching a tipping point for
recursive self-improvement. GPU kernel optimization lets LLMs improve software
governing their own execution
efficiency~\cite{Chen:2018, Cheng:2026}.

LLMs can now generate functionally correct kernels for diverse
workloads~\cite{Hu:2026, Dai:2026, Zhang:2025, Su:2025, Dong:2026a, Liao:2026,
Ouyang:2025, Wang:2025, Baronio:2026, Cao:2026, Dong:2026}, but their
performance remains unsuitable for production. With GPT-5.6 Sol on AMD MI300X,
existing agentic harness systems produce GEMM kernels that are
$1.8$--$14.3\times$ slower than hipBLASLt~\cite{AMD:2026}, AMD's
state-of-the-art GEMM library. Closing this gap requires more than functional
correctness: agents must coordinate optimizations across the software-hardware
stack while preserving program semantics. For example, a high-performance GEMM
implementation combines matrix cores~\cite{Luo:2024}, software
pipelining~\cite{Lam:1988}, memory-hierarchy optimizations~\cite{Li:2023}, and
instruction scheduling~\cite{Huerta:2025}, many of which require low-level
assembly. Kernel-design agents~\cite{Hari:2026, Yang:2026} combine expert
knowledge bases with powerful models and iterate using unit-test outcomes and
performance profiles, yet they struggle to apply performance-critical, low-level
optimizations for two reasons. First, validating such transformations requires
reasoning about nontrivial properties, including memory
aliasing~\cite{Ramalingam:1994} and data races~\cite{Betts:2012}, at the
assembly level. This reasoning consumes many tokens and expands the context,
ultimately degrading model performance~\cite{Liu:2024}. Second, unit tests and
performance profiling only provide sparse, end-to-end feedback. These signals
reveal that a kernel is incorrect or slow but do not identify the responsible
transformation or ordering.

One alternative is to trade optimization control for compiler automation by
targeting a higher-level language such as Triton~\cite{Tillet:2019}, which
delegates many low-level decisions to a tensor compiler~\cite{Chen:2018, Guan:2025,
Ma:2020, Tillet:2019, Wang:2026, Wu:2025, Zheng:2020}. In our evaluation, the
ROCm Triton baseline narrows the gap to approximately $2\times$. However, this
baseline does not expose fine-grained control over instruction scheduling or
wavefront specialization,
preventing agents from closing the remaining gap.

In this paper, we present \system, an agentic framework for automating complex,
multilevel GPU kernel optimizations. \system uses \emph{data-flow invariants} to
validate transformations and provide guardrails as LLMs apply them.
These compile-time assertions specify relationships among data values that must
hold throughout kernel execution, even across multiple layers of optimization.
Data-flow invariants are an established concept in information-flow
security~\cite{Myers:1999, Yip:2009, Zeldovich:2006}. \system adapts them to GPU
kernel optimization. For example, a GEMM kernel tiles and stages matrices
$\mathbf{A}$ and $\mathbf{B}$ in shared memory before passing them to hardware
matrix fused multiply-add (MFMA) instructions, which expect the operands in a
particular swizzled layout. A
data-flow invariant can assert that these instructions always receive correctly
paired elements of $\mathbf{A}$ and $\mathbf{B}$, regardless of the
transformations applied. By making such requirements explicit, \system gives the
agent concrete correctness constraints for intrusive transformations such as
software pipelining and data-layout reorganization.

Three challenges arise in realizing this approach. First, the invariant
language must express global properties concisely in one place while producing
concrete violation reports that guide repairs, such as the thread holding a
mismatched element and the program point where the mismatch occurs. Second,
enforcing invariants must incur zero runtime overhead. Optimized kernels
saturate hardware units, tailor access patterns to the memory hierarchy, and
depend on precise instruction schedules. The analysis must therefore operate
entirely at compile time across tiling, staging, and layout transformations,
without perturbing the generated code. Third, application-specific invariant
templates must be instantiated automatically. The assignment of $\mathbf{A}$ and
$\mathbf{B}$ tiles to wavefronts, their staging in shared memory, and their eventual
consumption by MFMA instructions vary across GEMM kernels, so \system must
instantiate the corresponding invariants without per-kernel human intervention.

\system combines three ideas: \emph{tag functions},
\emph{static analysis}, and an \emph{in-context reinforcement learning (ICRL)
agent}.

First, \system introduces \emph{tag functions and tag assertions}, a concise
language for expressing data-flow invariants. It generates and optimizes kernels
in a tile-based, Pythonic DSL inspired by CuTe~\cite{NVIDIA:2026} and
TileLang~\cite{Wang:2026}. Tag functions attach symbolic values, such as logical
coordinates, to tensor elements and propagate them through control and data
flow. Tag assertions require tags at a use site either to match or to differ.
Conformity assertions encode alignment constraints, such as requiring paired
matrix-multiplication operands to carry compatible tags. Non-conformity
assertions encode separation constraints, such as ensuring that concurrent
producers do not overwrite the same shared-memory region.

Second, \system uses \emph{static analysis} to enforce these invariants entirely
at compile time and turn failures into actionable feedback for the LLM coding
agent. The tags are purely symbolic and never materialized at runtime.
The \system compiler tracks their propagation using the abstract
interpretation~\cite{Cousot:1977} over a layout algebra~\cite{NVIDIA:2026} and
leverages an SMT solver~\cite{Moura:2008} to solve the resulting integer
constraints. To remain tractable for heavily optimized kernels, the analysis
is path-insensitive. When an assertion fails, the compiler reports a concrete
assignment witnessing the violating abstract state, including the thread ID,
data element, and program point, thereby guiding the agent's next revision.

Third, \system uses \emph{ICRL}~\cite{Dong:2026a, Monea:2025} to adapt a planner
that proposes optimizations and their corresponding invariant templates. The
planner observes the kernel implementation, the compile-time invariant violations,
and the runtime profiles, retrieves applicable optimizations from a persistent
knowledge base, and proposes optimization instructions with their templates
(Figure~\ref{f:arch}). A lowering agent binds the selected instructions to code
and instantiates the templates as tag functions and assertions. Each invariant
specifies a property that its transformation must preserve. \system makes
planner prompts learnable, uses invariant violations and runtime performance as
reward signals, and applies text gradients~\cite{Yuksekgonul:2024} to improve
subsequent optimization plans.

These three ideas reinforce one another. Tag functions alone provide only
declarative specifications. Static analysis can check properties but cannot
repair violations, and an ICRL agent alone lacks the structured feedback needed
to navigate a vast optimization space. Together, they form a closed loop: tag
functions express correctness properties, static analysis checks them at
compile time, and concrete assignments witnessing abstract violations provide
dense feedback. This feedback
lets the agent reason about abstract correctness properties instead of repairing
implementations through trial and error. \system thereby applies complex,
multilevel optimizations more reliably while reducing the effort required to
preserve correctness. Discovering new algorithmic
optimizations~\cite{Hari:2026} and fully verifying implementation
correctness~\cite{Betts:2012, Chatterjee:2025} remain outside its scope.

We evaluate \system on three production-level kernel families using GPT-5.6 Sol, including
general matrix multiplications (GEMM)~\cite{AMD:2026}, flash
attention~\cite{dao2023flashattention}, and mixture-of-experts (MoE)
feed-forward networks~\cite{Shazeer:2017}. On AMD MI300X, these kernels account
for up to 90\% of GPU execution time during LLM
inference~\cite{GLM-5-Team:2026}. Kernels optimized by \system achieve 89--99\%
of the effective throughput of state-of-the-art hand-optimized
libraries~\cite{AMD:2025a, AMD:2026}. Across the evaluated workloads, \system
improves geometric-mean throughput by $1.62$--$1176\times$ over uncontaminated
agentic baselines. For flash attention, \system outperforms
KernelFalcon~\cite{Wang:2025} by $3\times$. Beyond these kernel families,
\system can also generalize to KernelBench~\cite{Ouyang:2025}: with in-context
DSL examples, it solves all 100 Level~1 and 88 Level~2 tasks within three
attempts.

Concretely, we make the following contributions:

\begin{itemize}
\item We introduce data-flow invariants for agentic GPU kernel optimization,
improving the reliability and performance of generated kernels.
\item We design a tile-based DSL with tag functions and assertions, together
with a static analysis that validates invariants at compile time with zero
runtime overhead.
\item We build an agentic optimization harness with an ICRL planner that
proposes optimization candidates, a selector that constructs plans, and a
lowering agent that implements them and instantiates their invariants in the
\system DSL.
\item We evaluate \system on GEMM, flash attention, MoE, and KernelBench,
demonstrating substantial gains over existing agentic baselines.
\end{itemize}

% !Tex Root = ./paper.tex

\section{Motivating Example}
\label{s:background}

\begin{figure*}[ht]
\begin{minipage}{\textwidth}
\kern0pt
\hspace{0.015\textwidth}%
\begin{minipage}[t]{0.55\textwidth}
\kern0pt
\renewcommand{\ttdefault}{zi4n}
% ,basicstyle=\ttfamily\fontsize{7.8pt}{8.5pt}\selectfont
\begin{lstlisting}[style=substrate,escapeinside={(*}{*)},morekeywords={concat},numbers=left,numberstyle=\tiny\color{gray},numbersep=3pt,breaklines=false,breakatwhitespace=false,basicstyle=\ttfamily\footnotesize,lineskip=-0.75pt]
def attn(d:const,gqa:const,Q:Tensor((sq,h(*\textsubscript{q}*),d),bf16),K:Tensor((sq,h(*\textsubscript{kv}*),d),bf16),
  V:Tensor((sq,h(*\textsubscript{kv}*),d),bf16),O:Tensor((sq,h(*\textsubscript{q}*),d),bf16), (*$\ldots$*)):
  # Assign tags to QKV 
  T(*\textsubscript{Q}*),T(*\textsubscript{K}*),T(*\textsubscript{V}*) = Q[sq,h(*\textsubscript{q}*),d](*$\rightarrow$*)(h(*\textsubscript{q}*)/gqa,d), K[sq,h(*\textsubscript{kv}*),d](*$\rightarrow$*)(h(*\textsubscript{kv}*),d), V[sq,h(*\textsubscript{kv}*),d](*$\rightarrow$*)(h(*\textsubscript{kv}*),sq)
  # Vectorize and tile memory access for MFMA
  idx(*\textsubscript{q}*),idx(*\textsubscript{h}*),tid = blockIdx.x, blockIdx.y, threadIdx.x
  wid,wtid,nw,rO = tid/64,tid%64,(wid&2)*2+wid/4*2+wid%2,make_local((4,),u256)
  gQ,rQ = Q.view((sq/Br,Br,h(*\textsubscript{q}*),d/8),u128), make_local((8,),u128)
  gK,sK = K.view((sq/Bc,Bc,h(*\textsubscript{kv}*),d/8),u128), make_shared((512,2),u128)
  gV,sV = V.view((sq/Bc,Bc,h(*\textsubscript{kv}*),d/2),u32), make_shared((2,2,4,4,32),u64)
  (*$\widetilde{\texttt{sK}}$*),(*$\widetilde{\texttt{sV}}$*) = sK.view((2,32,8,2),u128), sV.view((2,2,4,4,16),u128)
  rS,(*$\widetilde{\texttt{rS}}$*),tU = make_local((2,),u256),rS.view((4,8),bf16),make_local((2,4,),u32)
  (*$\forall i \in [0,8)$*): rQ[i] = gQ[idx(*\textsubscript{q}*),wid*32+wtid%32,idx(*\textsubscript{h}*),i*2+wtid/32]
  for i in range(sq/Bc):
    for j in range(2):
      sK[tid,j] = gK[i,j*32+tid/16,idx(*\textsubscript{h}*)/gqa,tid%16]
      # Vectorize V(*\color{green!50!black}\textsuperscript{T}*) via bit rearrangements 
      (*$\forall k \in [0,4)$*): (*\colorbox{blue!10!white}{\tt tU[j,k] = gV[i,j*32+wid*4+k,idx\textsubscript{h}/gqa,wtid]}*)
      (*\colorbox{green!10!white}{\tt sV[0,j,wtid/16,nw\%4,(wtid\%16)*2+nw/4] = ${\color{blue!80!black}\texttt{concat}}^{3}_{k=0}$(tU[j,k]\textsubscript{lo})}*)
      (*\colorbox{red!10!white}{\tt sV[1,j,wtid/16,nw\%4,(wtid\%16)*2+nw/4] = ${\color{blue!80!black}\texttt{concat}}^{3}_{k=0}$(tU[j,k]\textsubscript{hi})}*)
    syncthreads()
    T(*\textsubscript{$\widetilde{\texttt{rS}}$}*) = (*$\widetilde{\texttt{rS}}$*)[x,y](*$\rightarrow$*)((*$\frac{\text{idx}\textsubscript{h}}{\text{gqa}}$*),i*Bc+(*$\frac{\text{x}}{2}$*)*32+x%2*16+(*$\frac{\text{y}}{4}$*)*8+(*$\frac{\text{wtid}}{32}$*)*4+y%4)
    for j in range(16):
      tK,tQ = (*$\widetilde{\texttt{sK}}$*)[j/8,wtid%32,j%8,wtid/32], rQ[j%8]
      (*$\widetilde{\texttt{tK}}$*),(*$\widetilde{\texttt{tQ}}$*) = tK.view((8,),bf16), tQ.view((8,),bf16)
      assert tag((*$\widetilde{\texttt{tK}}$*)[(*$\ldots$*)]) == tag((*$\widetilde{\texttt{tQ}}$*)[(*$\ldots$*)]) # (*\color{green!50!black}$\widetilde{\texttt{tQ}}$*) and (*\color{green!50!black}$\widetilde{\texttt{tK}}$*) should match
      rS[j/8] = matmul((*$\widetilde{\texttt{tK}}$*),(*$\widetilde{\texttt{tQ}}$*),(*$\ldots$*)) # FP32 accumulate and pack S=QK(*\color{green!50!black}\textsuperscript{T}*)
    (*$\ldots$*) # Online softmax: overwrite rS with packed P and rescale rO
    for j in range(16): # PV, V is transposed due to MFMA
      (*\colorbox{gray!20!white}{\tt tV = $\widetilde{\texttt{sV}}$[wtid\%2,j/8,j\%4,j\%8/4*2+wtid/32,(wtid/2)\%16]}*)
      (*$\widetilde{\texttt{tV}}$*) = tV.view((8,),bf16)
      assert tag((*$\widetilde{\texttt{tV}}$*)[(*$\ldots$*)]) == tag((*$\widetilde{\texttt{rS}}$*)[(*$\ldots$*)]) # (*\color{green!50!black}$\widetilde{\texttt{tV}}$*) and (*\color{green!50!black}$\widetilde{\texttt{rS}}$*) should match
      rO[j%4] = matmul((*$\widetilde{\texttt{tV}}$*),(*$\widetilde{\texttt{rS}}$*)[j/4],(*$\ldots$*)) # FP32 accumulate and pack O
    syncthreads()
  (*$\ldots$*) # Normalize, pack, and store O
\end{lstlisting}
\end{minipage}%
\begin{minipage}[t]{0.435\textwidth}
\kern0pt
\begin{minipage}[t]{\textwidth}
\centering
\usetikzlibrary{positioning}
\begin{tikzpicture}[
  x=1cm,
  y=1cm,
  font=\footnotesize,
  >=Latex,
  flow/.style={draw, line width=0.45pt, align=center, fill=white,
    inner xsep=3pt, inner ysep=2pt},
  arrow/.style={->, line width=0.45pt},
  label/.style={align=center, fill=white, inner sep=1pt}
]

% Inputs and initial kernel.
\node[flow, minimum width=2.88cm, minimum height=0.60cm]
  (description) {Kernel description};
\node[flow, minimum width=2.88cm, minimum height=0.60cm,
  right=0.43cm of description]
  (hardware) {HW specifications};
\coordinate (inputs-south) at
  ($(description.south)!0.5!(hardware.south)$);
\node[flow, minimum width=6.18cm, minimum height=0.60cm,
  below=0.34cm of inputs-south]
  (initial)
  {Initial kernel};
\draw[arrow] (description.south) -- (description.south |- initial.north);
\draw[arrow] (hardware.south) -- (hardware.south |- initial.north);

% Agentic optimization loop.
\node[flow, minimum width=3.12cm, minimum height=0.72cm,
  below=0.57cm of initial, xshift=-0.46cm]
  (planner)
  {Learnable Planner\\[-1pt](via ICRL)};
\node[flow, minimum width=3.12cm, minimum height=0.72cm,
  below=0.80cm of planner]
  (selector) {Optimization Selector};
\node[flow, minimum width=3.12cm, minimum height=0.76cm,
  below=0.82cm of selector]
  (lowering) {Lowering Agent};
\node[flow, minimum width=3.12cm, minimum height=0.74cm,
  below=0.72cm of lowering]
  (validation)
  {Invariant Validation\\[-1pt]Testing \& Profiling};

\node[draw=gray!70, dashed, line width=0.55pt,
  fit={($(planner.north west)+(-1.08,0.35)$)
       ($(validation.south east)+(0.77,-0.50)$)},
  inner sep=0pt] (harness) {};
\node[anchor=south east, inner sep=2pt] at (harness.south east)
  {Agentic Harness};

\coordinate (flow-axis) at ($(planner.south)+(-0.20,0)$);
\draw[arrow] (flow-axis |- initial.south) --
  (flow-axis |- planner.north);
\draw[arrow] (flow-axis |- planner.south) --
  node[label, xshift=1.12cm]
  {\itshape Optimization\\[-2pt]\itshape candidates}
  (flow-axis |- selector.north);
\draw[arrow] (flow-axis |- selector.south) --
  node[label, xshift=1.17cm]
  {\itshape Instructions +\\[-2pt]\itshape invariant templates}
  (flow-axis |- lowering.north);
\draw[arrow] (flow-axis |- lowering.south) --
  node[label, xshift=1.12cm]
  {\itshape \system DSL}
  (flow-axis |- validation.north);

% Persistent and runtime resources.
\node[flow, rotate=-90, anchor=north west,
  minimum width=2.90cm, minimum height=0.72cm]
  (knowledge) at (initial.east |- harness.north) {Knowledge Base};
\node[flow, rotate=-90, anchor=north east,
  minimum width=3.10cm, minimum height=0.72cm]
  (runtime) at (initial.east |- harness.south) {Runtime Libraries};
\draw[arrow] (knowledge.south |- planner.east) -- (planner.east);
\draw[arrow] (runtime.south |- lowering.east) -- (lowering.east);

% Feedback and output.
\draw[arrow] (validation.west) -- ++(-0.48,0)
  coordinate (feedback-west) |- (planner.west);
\node[label, rotate=90] at
  ($(feedback-west)!0.5!(feedback-west |- planner.west)$)
  {Compile-time \& runtime feedback};
\node[flow, minimum width=6.18cm, minimum height=0.60cm,
  below=0.70cm of validation, xshift=0.46cm]
  (output)
  {Optimized kernel};
\draw[arrow] (flow-axis |- validation.south) --
  (flow-axis |- output.north);

% Preserve the original figure's compact portrait footprint.
\path[use as bounding box]
  ($(output.south west)+(-0.035cm,-0.01cm)$) rectangle
  ($(hardware.north east)+(0.03cm,0)$);
\end{tikzpicture}
\end{minipage}

\begin{minipage}[t]{\textwidth}
\definecolor{Color0}{HTML}{A6CE33}
\definecolor{Color1}{HTML}{1F78B4}
\definecolor{Color2}{HTML}{B2DF8A}
\definecolor{Color3}{HTML}{33A02C}
\definecolor{Color4}{HTML}{FB9A99}
\definecolor{Color5}{HTML}{E31A1C}
\definecolor{Color6}{HTML}{FDBF6F}
\definecolor{Color7}{HTML}{FF7F00}
\kern0.01in
\begin{tikzpicture}[tag/.style={inner sep=0pt, minimum size=7pt},font=\footnotesize]
\fill[blue!10!white] (-0.1, 0.4*7.5) rectangle (0.5, 1.4);

\foreach \i [evaluate=\i as \y using (7-\i)*0.4] in {0,1,2,3} {
  \node[tag,circle,fill=Color\i] at (0,\y) {};
  \node[tag,circle,fill=Color\i] at (0.4,\y) {};
  \node[tag,circle,fill=Color\i] at (0.4*2,\y) {};
  \node[tag,circle,fill=Color\i] at (0.4*3,\y) {};
}
\node at (0,3*0.4+0.05) {$\vdots$};
\node at (0.4*4, 0.4*7) {r\textsubscript{0}};
\node at (0.4*4, 0.4*2) {r\textsubscript{8}};

\foreach \i [evaluate=\i as \y using (7-1-\i)*0.4] in {4,5,6,7} {
  \node[tag,circle,fill=Color\i] at (0,\y) {};
  \node[tag,circle,fill=Color\i] at (0.4,\y) {};
  \node[tag,circle,fill=Color\i] at (0.4*2,\y) {};
  \node[tag,circle,fill=Color\i] at (0.4*3,\y) {};
}

\draw[->] (-0.4, 0.4 * 8) -- (-0.4, -0.4) node[midway, below, sloped] {\texttt{sq}};
\draw[->] (-0.4, 0.4 * 8) -- (1.2, 0.4 * 8) node[midway, above] {\texttt{d}};
\node (gmem) at (0.4, 0.4*8 + 0.5) {Global memory};

\fill[green!10!white] (3-0.2, 0.4*9.5) rectangle (3-0.2+0.4*4, 0.4 * 8.5);
\fill[red!10!white] (3-0.2, 0.4*7.5-0.1) rectangle (3-0.2+0.4*4, 0.4 * 6.5-0.1);

\foreach \i [evaluate=\i as \x using 3+\i*0.4] in {0,1,2,3,4,5,6,7} {
  \node[tag,circle,fill=Color\i] at (\x, 0.4*9) {};
  \node[tag,circle,fill=Color\i] at (\x, 0.4*7-0.1) {};
  \node[tag,circle,fill=Color\i] at (\x, 0.4*8) {};
  \node[tag,circle,fill=Color\i] at (\x,0.4*6-0.1) {};
}
\draw [decorate, decoration={brace, amplitude=5pt, mirror, raise=2pt}] 
    (3, 0.4*6-0.2) -- (3+3*0.4, 0.4*6-0.2) 
    node [midway, below=7pt] {wid=0};
\draw [decorate, decoration={brace, amplitude=5pt, mirror, raise=2pt}] 
    (3+0.4*4, 0.4*6-0.2) -- (3+7*0.4, 0.4*6-0.2) 
    node [midway, below=7pt] {wid=2};
\node at (3 - 0.4, 0.4*8.5) {lo};  
\node at (3 - 0.4, 0.4*6.5) {hi};  
\node[rotate=270, text width=1.5cm, align=center] at (3.2 + 8.2 * 0.4, 0.4*7.5)
(shm) {Shared \\ memory};

\begin{scope}[shift={(0,-0.4)}]
\fill[gray!20!white] (3-0.2, 0.4*3.5) rectangle (3-0.2+0.4*8, 0.4 * 2.5);
\foreach \i [evaluate=\i as \x using 3+\i*0.4] in {0,1,2,3,4,5,6,7} {
  \node[tag,circle,fill=Color\i] at (\x, 0.4*3) {};
  \node[tag,circle,fill=Color\i] at (\x, 0.4*2) {};
  \node[tag,circle,fill=Color\i] at (\x, 0.4) {};
  \node[tag,circle,fill=Color\i] at (\x,0) {};
}

\node[rotate=270, text width=1.5cm, align=center] at (3.2 + 8.2 * 0.4, 0.4*1.5)
(shm) {Registers};
\draw[->] (3-0.4, 0.4 * 3.5) -- (3-0.4, 0) node[midway, below, sloped] {tid};
\end{scope}

\end{tikzpicture}
\end{minipage}
\end{minipage}
\end{minipage}
%\vspace{-0.5em}
\caption{Overview of \system. Left: Simplified DSL implementation of flash
attention ($d{=}128$, $B_r{=}256$, $B_c{=}64$, 512 threads). Top right:
\system's workflow for agentic optimization. Bottom right: Tag propagation for
$V$ across memory levels. Each color represents a unique sequence-coordinate
tag, and background shading links code regions to memory access patterns.
r\textsubscript{0} and r\textsubscript{8} represent row 0 and row 8. 
}
\Description{A three-part overview of Ave. Pseudocode on the left implements
flash attention in the Ave DSL. A flowchart on the upper right shows the
planner, optimization selector, lowering agent, and validation loop. A diagram
on the lower right traces sequence-coordinate tags for V through global memory,
shared memory, and registers for a fixed attention head.}
\label{f:arch}
\floatrule
\end{figure*}

We use flash attention~\cite{dao2023flashattention} on the AMD MI300X GPU as a
running example to illustrate why high-performance kernel development demands
tightly coupled algorithmic and hardware-level reasoning. Together with GEMM and
MoE layers, attention accounts for up to 90\% of GPU execution time in LLM
serving~\cite{GLM-5-Team:2026}, making its optimization critical.

The standard attention mechanism operates on three input matrices
$\mathbf{Q}, \mathbf{K}, \mathbf{V} \in \mathbb{R}^{N\times d}$, where $N$ is the sequence length and $d$
is the head dimension. It computes $\mathbf{S}=\mathbf{QK}^T$, applies softmax
row-wise to obtain $\mathbf{P}=\text{softmax}(\mathbf{S})$, and produces
$\mathbf{O}=\mathbf{P}\mathbf{V}$. A na\"{\i}ve implementation materializes the
$N \times N$ score matrix $\mathbf{S}$,
requiring $\mathcal{O}(N^2)$ memory, prohibitive for long contexts (e.g., $N = 256\text{K}$). Flash attention avoids this by tiling:
it loads a $B_r \times d$ block of $\mathbf{Q}$ into registers, then streams
through $\mathbf{K}$ and $\mathbf{V}$ in $B_c \times d$ blocks staged in shared
memory, accumulating into $\mathbf{O}$ via online softmax~\cite{milakov2018online}
to achieve $\mathcal{O}(N)$ memory. The tile sizes $B_r$ and $B_c$ are chosen so
the working set fits in registers and shared memory.

\inlinesection{GPU Architecture.}
\system targets the AMD MI300X GPU, which implements the CDNA3
architecture~\cite{AMD:2025}. Like other modern GPUs, the MI300X follows a
Single Instruction, Multiple Threads (SIMT) design. Each Compute Unit (CU)
contains in-order execution cores, a shared register file, and Local Data Share
(LDS), or shared memory, and is analogous to an NVIDIA streaming multiprocessor.
A 64-thread \emph{wavefront} (AMD's warp) executes in lockstep on a 16-lane
vector unit. Several wavefronts form a \emph{workgroup} sharing LDS. MI300X also
provides MFMA matrix cores, 2-wide vector fused multiply-add instructions, and
Vector Accumulation Registers (AGPRs) that expand the available register pool.
These features require explicit compiler support.

Achieving peak performance on a GPU requires exploiting this architecture along
three axes. First, kernels balance per-workgroup register and shared-memory use
against occupancy. Second, MI300X's arithmetic-to-bandwidth ratio of $\sim$250
demands vectorized accesses, shared-memory reuse, and software pipelining to
overlap computation with transfers. Third, instruction scheduling and register
allocation matter at the assembly level: poor schedules stall in-order
pipelines, while excessive register use causes
spills~\cite{Quintao-Pereira:2008}.

\inlinesection{Optimizing Flash Attention on MI300X.}
Turning flash attention into a kernel reaching peak throughput requires a
cascade of hardware-specific optimizations. Figure~\ref{f:arch} shows a
simplified \system DSL implementation with 512 threads, $d{=}128$, $B_r{=}256$,
and $B_c{=}64$, embodying these optimizations.
The pseudocode focuses on layouts and operand pairing, eliding the full pipeline
schedule and wavefront-group control flow.

The kernel executes two GEMMs $\mathbf{S}=\mathbf{Q}\mathbf{K}^T$ and
$\mathbf{O}= \mathbf{P}\mathbf{V}$ on matrix cores via MFMA
instructions. Each $32{\times}32{\times}8$ MFMA expects column-major input
operands $\mathbf{A}$ and $\mathbf{B}$ and accumulator $\mathbf{C}$ in
hardware-specific swizzled layouts. Each thread contributes 4 elements from
each input. Adapting row-major attention to these layouts requires operand reordering
and transposition. Reaching peak throughput also requires shared-memory
reuse, software pipelining to overlap global memory loads with MFMA computation,
partitioning data-flow across disjoint wavefront groups, and careful instruction
scheduling to keep the in-order cores busy without pipeline stalls while staying
within the register budget to avoid spills. For example, AITER's handwritten
assembly kernel uses 11 pipeline stages across two wavefront groups.
Section~\ref{s:dsl} details how the DSL expresses these optimizations.

The data shuffling and layout transformations are \textit{error-prone}:
mismatches between expected and actual data layouts cause correctness bugs.
Ensuring correctness of these global properties is time-consuming when drafting
and debugging optimizations. \system introduces tag functions and data-flow
invariants, expressed as {\tt assert} statements in Figure~\ref{f:arch}, to
check that values are correctly paired at each use site. Section~\ref{s:dsl}
describes these mechanisms in detail.

% !Tex Root = ./paper.tex

\section{Overview of \system}
\label{s:overview}

\system combines three components to generate and optimize GPU kernels:
(\underline{i})~a tile-based, Pythonic DSL with tag functions and assertions
(\S\ref{s:dsl}), (\underline{ii})~compile-time invariant analysis with zero
runtime cost (\S\ref{s:analysis}), and (\underline{iii})~an agentic harness
guided by invariant feedback (\S\ref{s:harness}). The implementation primarily
targets AMD MI300X and generalizes to AMD MI350X and NVIDIA H200
(\S\ref{sec:exp_q3}).

\inlinesection{\system DSL.} The Pythonic, tile-based DSL resembles CuTe and
Triton and adds tag functions and assertions for data-flow invariants. It
exposes hardware instructions and compiler policies while abstracting complex
memory layout encodings as tiles (\S\ref{s:dsl}).

\inlinesectiontighter{Invariant Validation.} \system validates data-flow invariants
at compile time with zero runtime overhead. The compiler lowers the DSL to
MLIR~\cite{Lattner:2021} and propagates tags across assignments and shared
memory using flow-sensitive, path-insensitive analysis. Path-insensitive merging
keeps solving tractable. An SMT solver checks tag assertions and returns concrete
assignments that witness violations and guide the agent
~\cite{Moura:2008} (\S\ref{s:analysis}).

\inlinesectiontighter{Agentic Optimization Harness.} \system adopts an agentic workflow to
iteratively optimize kernels, as shown at the top right of
Figure~\ref{f:arch}. Starting from a DSL kernel and unit tests, a planner
retrieves optimization knowledge, a selector orders compatible proposals, and a
lowering agent implements them. Validation checks invariants and tests before
profiling. Invariant violations and performance update the benchmark-local
planner through ICRL (\S\ref{s:harness}).

% !Tex Root = ./paper.tex

\section{The \system DSL}
\label{s:dsl}

\inlinesection{Layouts and Tiles.} The \system DSL models all memory
accesses via {\em tiles}~\cite{NVIDIA:2026, Li:2026, Wang:2026, Ding:2025}. A
tile is a memory region paired with a layout function~\cite{Carlisle:2026} that
maps multi-dimensional logical coordinates to one-dimensional physical offsets.
A layout function $\mathcal{L}_{(\mathbf{s}, \mathbf{t})}$ is parameterized by
two tuples: shapes $\mathbf{s}=(s_1, \ldots, s_n)$ and strides $\mathbf{t}=(t_1,
\ldots, t_n)$. It maps an $n$-element coordinate $\mathbf{c} = (c_1, \ldots,
c_n)$ to a linear index. Following the CuTe layout algebra, when all elements of
$\mathbf{s}$ and $\mathbf{t}$ are integers, the layout computes
\[ \mathcal{L}(c_1, \ldots, c_n) = \sum_{i=1}^n c_i t_i \]

\noindent For a contiguous row-major tensor, $t_i=\prod_{j=i+1}^{n}s_j$: each stride is
the product of the dimensions that follow it, and $t_n=1$. For brevity, these
strides are omitted. For example, the {\tt
Q:Tensor((sq,h\textsubscript{q},d),bf16)} in Figure~\ref{f:arch} has the layout
\[ \mathcal{L}^{\mathbf{Q}}(c_1,c_2,c_3) =c_1\texttt{h\textsubscript{q}}\texttt{d}+c_2\texttt{d}+c_3. \]
\noindent The subscript operator in Figure~\ref{f:arch} maps coordinates to a
linear index and accesses the corresponding memory location.

Following CuTe, a nested layout represents its shape, stride,
and natural coordinate as congruent hierarchical tuples: every coordinate leaf
pairs with the stride leaf at the same position. A scalar coordinate for a
nested mode is first converted to this hierarchical form by mixed-radix
division and remainder using the mode's shape. The layout recursively sums
the coordinate--stride products. For example, for the first nested coordinate
$\mathbf{c}_1=(c_{1,1},c_{1,2})$, with shape $(s_{1,1},s_{1,2})$ and stride
$(t_{1,1},t_{1,2})$, the corresponding sub-layout computes
$\mathcal{L}_1(\mathbf{c}_1)=c_{1,1}t_{1,1}+c_{1,2}t_{1,2}$.
Nested layouts primarily represent hardware-swizzled
layouts used by tensor cores and matrix cores. Layout algebras further define
composition, logical division, and logical projection over layout functions to
represent operations such as data shuffling and dispatching disjoint work to
warps (wavefronts on AMD) and threads. We refer the reader
to~\cite{Carlisle:2026} for details.

DSL programs use {\tt make\_local()} and {\tt make\_shared()} to allocate a tile
backed by registers and shared memory. DSL programs may use the {\tt view()}
function to reinterpret a tile under a different logical layout. For example,
line 8 of Figure~\ref{f:arch} derives {\tt gQ} to logically partition {\tt Q} by
$B_r$ and access data in 16-byte chunks:
\begin{lstlisting}[style=substrate,escapeinside={(*}{*)},numbers=none,
  basicstyle=\small\ttfamily\centering,aboveskip=3pt,belowskip=3pt]
gQ = Q.view((sq/Br,Br,h(*\textsubscript{q}*),d/8),u128)
\end{lstlisting}
\noindent Vectorizing loads in 16-byte chunks is a common optimization to fully
utilize GPU memory bandwidth. However, \system requires tiles to express all
memory accesses and restricts layout operations. For example, {\tt view()}
requires source and destination layouts to cover identical memory sizes. These
constraints simplify program analysis and eliminate the need for full-scale
alias analysis.

Continuing with the flash attention example in Figure~\ref{f:arch}, the kernel
maps work across 8 wavefronts. Each wavefront owns 32 rows of $\mathbf{Q}$ and
produces the corresponding rows of $\mathbf{O}$, while all wavefronts share
$B_c{\times}d$ tiles of $\mathbf{K}$ and $\mathbf{V}$ in shared memory. MFMA
expects column-major input operands $\mathbf{A}$ and $\mathbf{B}$ and
accumulator $\mathbf{C}$. In attention, $\mathbf{S}$ must be row-major because
$\mathbf{P}{=}\text{softmax}(\mathbf{S})$ is computed row-wise, and $\mathbf{O}$
must also be row-major. The kernel therefore swaps $\mathbf{Q}$ and $\mathbf{K}$
for $\mathbf{QK}^{T}$ and transposes $\mathbf{V}$ during staging for
$\mathbf{PV}$. The {\tt matmul} helper combines two MFMAs along the reduction
dimension to perform a $32{\times}32{\times}16$ operation per wavefront. Each
thread holds 8 BF16 elements from each input and 16 FP32 accumulator elements.

In Figure~\ref{f:arch}, {\tt matmul} accumulates $\mathbf{S}$ in FP32 and packs
it into the BF16 {\tt rS} tile. The elided online softmax computes $\mathbf{P}$
row-wise, overwrites {\tt rS} with the packed probabilities, and rescales
{\tt rO}. To transpose $\mathbf{V}$, each
thread loads a $4{\times}2$ sub-tile into {\tt tU} (line~18) and scatters its low
and high halves into transposed positions in {\tt sV} (lines~19--20). This
organization uses 4-byte global loads, 8-byte shared-memory stores, and 16-byte
swizzled shared-memory loads that feed the $\mathbf{PV}$ loop directly
(lines~29--33, highlighted in gray). {\tt matmul} accumulates the output in FP32
and packs it into {\tt rO}. After all sequence tiles, the kernel normalizes,
packs, and stores $\mathbf{O}$. The final steps and lane-to-output mapping are
elided in the figure.

\inlinesection{Tag Functions and Assertions.} A tag function annotates tensor
elements with symbolic tags: tuples of constants, symbolic expressions of
coordinates, or the special symbols $\bot$ and $\top$. Tags propagate through
assignments.

For example, {\tt T\textsubscript{Q}}, {\tt T\textsubscript{K}}, and {\tt
T\textsubscript{V}} in Figure~\ref{f:arch} assign tags to elements of {\tt Q},
{\tt K}, and {\tt V} based on their coordinates (the type \texttt{bf16} is
omitted for brevity):
{\small
\setlength{\abovedisplayskip}{3pt}
\setlength{\belowdisplayskip}{3pt}
\setlength{\abovedisplayshortskip}{3pt}
\setlength{\belowdisplayshortskip}{3pt}
\begin{displaymath}
\setlength{\arraycolsep}{0.2ex}
\begin{array}{lcl}
\text{\tt T\textsubscript{Q} = Q[sq,h\textsubscript{q},d]} &\rightarrow& \text{\tt (h\textsubscript{q}/gqa, d)} \\
\text{\tt T\textsubscript{K} = K[sq,h\textsubscript{kv},d]} &\rightarrow& \text{\tt (h\textsubscript{kv}, d)} \\
\text{\tt T\textsubscript{V} = V[sq,h\textsubscript{kv},d]} &\rightarrow& \text{\tt (h\textsubscript{kv}, sq)}
\end{array}
\end{displaymath}
}

\noindent Tag assertions require conformity or
non-conformity between tags. The assertion on line~26 in Figure~\ref{f:arch}
requires matching tags for elements of {\tt Q} and {\tt K}:
\begin{lstlisting}[style=substrate,escapeinside={(*}{*)},numbers=none,
  basicstyle=\small\ttfamily\centering,aboveskip=3pt,belowskip=3pt]
assert tag((*$\widetilde{\texttt{tK}}[\ldots]$*)) == tag((*$\widetilde{\texttt{tQ}}[\ldots]$*))
\end{lstlisting}
\noindent Computing $\mathbf{S}=\mathbf{Q}\mathbf{K}^T$ requires each wavefront to
multiply its 32 rows of $\mathbf{Q}$ with two $32{\times}128$ blocks of
$\mathbf{K}$ per iteration. The tag functions {\tt T\textsubscript{Q}} and
{\tt T\textsubscript{K}} return the attention-head and reduction-dimension
indices, so their tags match only when the operands belong to the same head and
represent the same position along the reduction dimension. A passing assertion confirms
that the implementation preserves this operand-pairing requirement across
optimizations such as shared-memory staging or software pipelining. Similarly,
the tag functions {\tt T\textsubscript{V}} and {\tt
T\textsubscript{$\widetilde{\texttt{rS}}$}} return the attention-head and
reduction-dimension indices. The assertion on line~32 therefore checks that the
operands of $\mathbf{O}=\mathbf{P}\mathbf{V}$ belong to the same head and
represent the same position along their shared reduction dimension, without
equating their independent output dimensions. Figure~\ref{f:arch} (bottom
right) visualizes {\tt V}'s sequence-coordinate tags as data propagate from
global memory to shared memory and registers.

\inlinesection{Compiler and Hardware Primitives.} The DSL exposes
compiler and hardware primitives that enable agents to explore
low-level optimizations. The DSL operates at the thread
level: kernels manage thread indices and synchronization (e.g.,
\texttt{syncthreads()}) directly, enabling the topology-aware dispatch
optimizations~\cite{Wang:2026}.

\system also exposes the eager materializations and compiler scheduling barriers for
tuning instruction scheduling in the backend. These compiler barriers constrain
instruction placement around runtime synchronization primitives such as
\texttt{s\_barrier}. Unlike \texttt{s\_barrier}, they emit no synchronization
instruction. These interfaces provide LLMs explicit control over hardware
features that are critical for reaching peak performance on
GPUs~\cite{Hu:2025,Triton:2026}.

Additionally, \system exposes vectorized instructions for explicit use in the
DSL. For example, the AMD MI300X issues one 2-way vectorized FMA per cycle on
packed registers, doubling scalar FMA throughput. The AMD GPU backend can emit
these instructions, but suboptimal register allocation often prevents it, so
explicit vectorization in the DSL sidesteps this limitation. \system further
exposes instructions such as \texttt{buffer\_load\_dwordx4} for branchless
memory access with hardware out-of-bounds (OOB) protection, and allows selecting the
accumulator register class (AGPR) directly in buffer load and MFMA
instructions, practically doubling available vector registers over
off-the-shelf toolchains.

% !Tex Root = ./paper.tex
\section{Validating Data-Flow Invariants}
\label{s:analysis}

\system validates data-flow invariants by tracking tags through control and data
flow at compile time. The compiler lowers DSL programs to MLIR and performs a
flow-sensitive, path-insensitive analysis that propagates tags through
assignments and shared-memory accesses. At control-flow joins, \system merges
incoming tags in a lattice with the partial order $\bot < t < \top$ for every
concrete tag $t$:
\[
\mathrm{merge}(t_1, t_2) = \begin{cases}
t_1 & \text{if } t_2 \le t_1 \\
t_2 & \text{if } t_1 < t_2 \\
\top & \text{otherwise}
\end{cases}
\]
Constants carry $\bot$. Consider the guarded selection
\texttt{rQ = 0 if isOOB(...) else Q[...]}, where \texttt{isOOB} prevents an
out-of-bounds access. The zero-valued branch contributes $\bot$, so
\texttt{rQ} retains the tag of \texttt{Q[...]} because
$\mathrm{merge}(\bot,t)=t$. Conversely, merging different concrete tags yields
$\top$, denoting non-conformity. Accordingly, \texttt{==} requires the merge to
be below $\top$, whereas \texttt{!=} requires $\top$: $\bot$ conforms with any
concrete tag, while $\top$ does not conform even with itself. \system can
reset shared-memory tags to $\bot$, permitting storage to be reused safely
across pipeline stages without disabling subsequent tag tracking.

The analysis tracks memory at byte granularity. This granularity covers the
evaluated byte-aligned accesses and supports packed storage for formats such as
\texttt{mxfp4} and \texttt{nvfp4}, although we do not evaluate FP4 kernels.
\system models global and shared memory as typed chunks and
automatically propagates tags across accesses of different widths, such as loads
vectorized into 16-byte chunks. To keep the analysis bounded, it does not model
the full heap or track tags written to global memory. The trade-off is that
\system cannot express invariants depending on heap-resident values. For
example, in MoE sorting, where global {\tt <expert,token>} tuples are grouped
by expert ID, \system cannot assert that each tuple reaches the correct bucket in
shared memory. Once data is in shared memory or registers, \system can assign
address-derived labels and track transformations such as swizzling. Multi-GPU
MoE kernels such as COMET~\cite{Zhang:2025a} exchange data via global
cross-GPU buffers, which \system does not currently track. Modeling these
buffers as explicit producer--consumer queues remains future work.

\system encodes propagated tags and tag assertions as satisfiability modulo
theories (SMT) constraints and uses the solver to reason about the resulting
layout algebra. The goal is to validate data-flow invariants, not to construct
full formal proofs. In particular, path-insensitive merging avoids enumerating
branch conditions and keeps analysis tractable for optimized kernels. When an
assertion fails, the solver returns a concrete assignment witnessing the
violating abstract state, including values for symbolic program variables such
as thread IDs, loop indices, and tensor coordinates.

Tags are purely \emph{compile-time constructs}: all tracking and validation
occur within the compiler and incur no runtime overhead. The current prototype
requires workgroup dimensions known at compile time, allowing \system to
model registers as a fixed-size array indexed by thread ID. It also unrolls
loops before analysis.

% !Tex Root = ./paper.tex

\section{Agentic Optimization Harness}
\label{s:harness}

The \system harness starts from a compile-and-test-passing DSL kernel and
retains the fastest candidate that passes automated invariant checks and unit
tests. In each round,
a planner proposes optimization candidates, a selector constructs an ordered
plan, and a lowering agent implements it. Only a measured speedup replaces the
current kernel.

\inlinesection{Persistent Expert Knowledge.} The knowledge base $\mathcal{K}$
contains auditable, expert-curated hardware facts, DSL references, and
optimization recipes. Recipes specify applicability, transformations, hardware
controls, and invariant templates. We curate entries from KernelWiki-style
documentation~\cite{KernelWiki:2026}, vendor examples, public pull requests,
and microbenchmarks. \textsc{Retrieve} filters them by the request, GPU, and
current kernel to produce $K_t$.

\inlinesection{Local Planner State.} For a request denoted by $q$, the fixed LLM
$\pi_\phi$ uses a benchmark-local prompt $\theta_{q,t}$ updated after each
optimization round through in-context reinforcement learning
(ICRL)~\cite{Dong:2026a,Monea:2025}. In round $t$, it observes
$x_t=\langle q,s^\star,K_t,F\rangle$: the best kernel $s^\star$, retrieved knowledge $K_t$, and
feedback history $F$. The initial history entry contains the initial kernel
$s_0$ and its profile. Later entries record candidates, validation feedback,
and profiles. The LLM produces ranked proposals $P_t$ specifying optimizations,
preconditions, target regions, invariant templates, and expected benefits.
\textsc{Select} orders compatible proposals into
$a_t=(a_{t,1},\ldots,a_{t,m_t})$.

\inlinesection{Lowering and Validation.} \textsc{Lower} invokes the lowering
agent to apply every optimization in $a_t$ to $s^\star$. The agent binds each
selected instruction to code, specializes its parameters, and materializes its invariant
templates as tag functions and assertions, returning $\hat{s}_t$.
Validation compiles this candidate, checks every generated invariant (the set
may be empty), and runs unit tests. \textsc{Validate} returns compilation status
$C_t$, source-level assertion sites $A_t$, failed sites $V_t$, test status
$U_t$, and record $f_t$. For a passing candidate, \textsc{Profile} returns its
latency and appends the runtime profile to $f_t$. Appending $f_t$ to $F$ gives
the next round its transformation and outcome.

\inlinesection{Reward Function.} Let $C_t\in\{0,1\}$ indicate successful
compilation and $U_t\in\{0,1\}$ indicate that all unit tests pass. Let $A_t$ be
the source-level assertion sites in $\hat{s}_t$, and let $V_t\subseteq A_t$ be
those with at least one failing check. Thus, compiler unrolling does not change
the number of assertions. If compilation succeeds and $A_t\ne\emptyset$, the
invariant score is $I_t=1-|V_t|/|A_t|$. Otherwise, $I_t=0$. If compilation
fails, invariant checking does not run, and we define $V_t=\emptyset$. Unit
tests run only when $C_t=1$ and $V_t=\emptyset$. If they do not run, $U_t=0$.
For a kernel that compiles, passes every invariant check, and passes all unit
tests, let $\hat{\ell}_t$ be its measured latency and $\ell^\star$ the latency
of the current best kernel. Runtime improvement is defined as:
\[
\Delta_t =
\begin{cases}
\displaystyle\frac{\ell^\star-\hat{\ell}_t}{\ell^\star},
 & C_t=1\ \land\ V_t=\emptyset\ \land\ U_t=1,\\[6pt]
0, & \text{otherwise}.
\end{cases}
\]
\noindent Thus, a positive $\Delta_t$ denotes lower latency, while a validated
slowdown receives a negative performance term. The complete reward is
$R_t=C_t+I_t+U_t+\Delta_t$. The replay record stores the planner input $x_t$,
candidate record $f_t$, validation outcomes $C_t$, $A_t$, $V_t$, and
$U_t$, performance term $\Delta_t$, and reward $R_t$. We denote this record by
$d_t$.

\begin{algorithm}[t]
\caption{ICRL-guided optimization.}
\label{alg:icrl}
\small\linespread{0.9}\selectfont
\KwIn{Query $q$, initial compile-and-test-passing kernel $s_0$, fixed knowledge
base $\mathcal{K}$, fixed LLM $\pi_\phi$, initial prompt $\theta_{q,0}$,
optimization rounds $N$}
$s^\star\leftarrow s_0$;
$(\ell^\star,f_{-1})\leftarrow
  \textsc{Profile}(s_0,\langle s_0\rangle)$;
$F\leftarrow\langle f_{-1}\rangle$\;

\For{$t=0,1,\ldots,N-1$}{
  $K_t\leftarrow\textsc{Retrieve}(\mathcal{K},q,s^\star)$\;
  $x_t\leftarrow\langle q,s^\star,K_t,F\rangle$\;
  $P_t\leftarrow\pi_\phi(x_t;\theta_{q,t})$\;
  $a_t\leftarrow\textsc{Select}(P_t)$\;
  $\hat{s}_t\leftarrow\textsc{Lower}(s^\star,a_t)$\;
  $f_t\leftarrow\langle a_t,\hat{s}_t\rangle$\;
  $(C_t,A_t,V_t,U_t,f_t)
    \leftarrow\textsc{Validate}(\hat{s}_t,f_t)$\;
  $(I_t,\Delta_t)\leftarrow(0,0)$\;
  \If{$C_t=1\land A_t\ne\emptyset$}{
    $I_t\leftarrow1-|V_t|/|A_t|$\;
  }
  \If{$C_t=1\land V_t=\emptyset\land U_t=1$}{
    $(\hat{\ell}_t,f_t)\leftarrow\textsc{Profile}(\hat{s}_t,f_t)$\;
    $\Delta_t\leftarrow
      (\ell^\star-\hat{\ell}_t)/\ell^\star$\;
  }
  $(R_t,F)\leftarrow
    (C_t+I_t+U_t+\Delta_t,F\mathbin{\|}\langle f_t\rangle)$\;
  $d_t\leftarrow
    \langle x_t,f_t,C_t,A_t,V_t,U_t,\Delta_t,R_t\rangle$\;
  \If{$\Delta_t>0$}{
    $(s^\star,\ell^\star)\leftarrow(\hat{s}_t,\hat{\ell}_t)$\;
  }
  $E_t\leftarrow\textsc{PolicyEval}(d_t)$\;
  $g_t\leftarrow\textsc{Analyze}(E_t)$\;
  $\theta_{q,t+1}\leftarrow
    \textsc{ParameterUpdate}(\theta_{q,t},g_t)$\;
}
\KwOut{Best kernel $s^\star$}
\end{algorithm}

\inlinesection{Text-Gradient Planner Update.} Algorithm~\ref{alg:icrl}
summarizes the planner update. After each optimization round,
\textsc{PolicyEval} summarizes $d_t$ as $E_t$, \textsc{Analyze} converts it
into a textual critique $g_t$, and \textsc{ParameterUpdate} applies a
constrained text-gradient edit to produce
$\theta_{q,t+1}$~\cite{Yuksekgonul:2024}. Neither the LLM weights nor
$\mathcal{K}$ change.

\inlinesection{Example.} We provide a concrete example to illustrate the process.
On AMD MI300X, hardware entries cover MFMA, buffer operations, data parallel
primitive (DPP) instructions, and LDS, vector general-purpose register (VGPR),
and AGPR limits. The omitted \texttt{matmul} helper in Figure~\ref{f:arch} uses
two $32{\times}32{\times}8$ BF16 MFMA instructions for a
$32{\times}32{\times}16$ operation. For row-major $QK^T$, it swaps $Q$ and $K$
as detailed in \S\ref{s:dsl}. Figure~\ref{f:arch}
keeps $Q$ in \texttt{rQ} and stages $K$ in \texttt{sK}, read through
$\widetilde{\texttt{sK}}$. The planner targets this region with a recipe whose
templates define the $Q/K$ tags and operand assertion. Here, $a_t$ contains
only this action. Lowering preserves existing assertions and adds this one with
the prompt:

\begin{quote}\small
Tile the scalar $QK^T$ dot products into $32{\times}32{\times}16$ accumulations,
using two $32{\times}32{\times}8$ BF16 MFMA instructions per accumulation. Pass
BF16 views of \texttt{tK} and \texttt{tQ} to \texttt{matmul}, which accumulates
in FP32 and packs the result into BF16 scores.
\end{quote}

The lowering agent materialized the tags and assertion and staged $K$ in
\texttt{sK}, but formed \texttt{tK} from the original $(512,2)$ layout. The
kernel compiled, but the tags of $\widetilde{\texttt{tK}}$ and
$\widetilde{\texttt{tQ}}$ disagreed at the matmul use site, so tests and
profiling did not run. Thus, $(C_t,U_t,\Delta_t)=(1,0,0)$, $I_t<1$, and
$R_t=1+I_t<2$. The validator added the following witness to $f_t$, which the
harness subsequently appended to $F$:

\begin{quote}\small
Assertion failed at \texttt{attn:26}:
$\operatorname{tag}(\widetilde{\texttt{tK}}[0])
\ne \operatorname{tag}(\widetilde{\texttt{tQ}}[0])$.\\
Witness: i=0, j=1, tid=0, wid=wtid=0,
idx\textsubscript{q}=idx\textsubscript{h}=0.
\end{quote}

The witness identified the mismatched MFMA operands. \textsc{PolicyEval}
summarized $d_t$, including the candidate code and witness, as $E_t$.
\textsc{Analyze} then traced the mismatch to the flattened LDS access and
produced the critique $g_t$.
\textsc{ParameterUpdate} applied $g_t$ to produce $\theta_{q,t+1}$. In the next
round, the planner proposed the following revision, which \textsc{Select}
placed at $a_{t+1,1}$:

\begin{quote}\small
Stage $K$ in \texttt{sK} with 16-byte cooperative loads,
synchronize, and reinterpret its $(512,2)$ layout as $(2,32,8,2)$
\texttt{u128}. For MFMA-loop index $j\in[0,16)$, let
$l=\texttt{wtid}$. Read \texttt{tK}
from $\widetilde{\texttt{sK}}$ at $(j/8,l\%32,j\%8,l/32)$, pair it with
$\texttt{tQ}=\texttt{rQ}[j\%8]$, and view both as 8 BF16 elements. Update
\texttt{rS[j/8]} with \texttt{matmul}, which accumulates these views in FP32
and packs the result into \texttt{rS[j/8]}.
\end{quote}

The revised binding reads $K$ through $\widetilde{\texttt{sK}}$ and leaves
\texttt{rQ} unchanged. It passed all checks and was profiled.

\section{Implementation}
\label{s:impl}

The \system prototype comprises approximately 22,000 lines of C++ and 7,000
lines of Python across the DSL compiler, program analysis, agentic harness, and
JIT infrastructure. We use Codex as the harness runtime~\cite{OpenAI:2025} and
configure the planner, lowering agent, \textsc{PolicyEval}, \textsc{Analyze},
and \textsc{ParameterUpdate} with the same selected model. \system retains the
current best kernel and summaries of prior candidates, validation outcomes, and
profiles. The harness periodically compacts the feedback history.  The
Python front-end parses DSL programs into an AST, which the C++ compiler lowers
to an extended MLIR dialect encoding layout algebras, then to LLVM IR and GPU
executables. We extend the LLVM~20 AMDGPU backend with explicit synchronization
barrier scheduling, register-class selection for buffer load and MFMA
instructions, and eager materialization through phantom uses in inline assembly.

Tag propagation and invariant checking use custom passes built on MLIR's pass
infrastructure. These passes traverse the IR and emit SMT constraints encoding
data-flow invariants, which Z3 solves~\cite{Moura:2008}.

The JIT layer follows Triton's model: generics propagate compile-time constants
into GPU kernels, compiled binaries are cached, and Python bindings expose kernel
invocation.

% !Tex Root = ./paper.tex
\section{Discussion}
\label{s:discussion}

\inlinesection{Limitations of Data-Flow Invariants.} Data-flow invariants
validate concrete transformations but do not propose algorithmic changes, such
as restructuring standard attention into tiled flash attention with online
softmax~\cite{dao2023flashattention}, or selecting search-space parameters such as
tile sizes and warp/wavefront specialization. The planner and knowledge base
handle these decisions. Applying invariants while the algorithm is still
changing could also constrain exploration. A full optimization can be separated
into two stages: algorithm selection and initial kernel generation occur first,
after which data-flow invariants guard low-level optimizations.

Full formal verification can establish stronger properties, but requires
substantial proof engineering~\cite{Betts:2012, Hawblitzel:2015, Klein:2009} and
produces implementation-specific proofs. \system instead checks shared-memory
layouts and operand pairing with tractable, flow-sensitive, path-insensitive
analysis. As Section~\ref{s:analysis} details, properties involving
global-memory writes, such as MoE routing and multi-GPU communication, require
new abstractions. The tensor memory (TMEM) extension discussed below illustrates
one such extension~\cite{NVIDIA:2026a}. Additionally, the analysis tracks memory
at byte granularity. It
treats two packed FP4 values within a byte as one unit, so permutations of
individual nibbles are invisible to data-flow invariants.

\inlinesection{Using More Tokens and Stronger Models.} As shown in
Section~\ref{s:eval}, stronger models often improve generated GPU kernels, but
higher token use alone does not close the gap with expert libraries: without
invariants, lowering agents can still exhaust their contexts through
trial-and-error repairs.

Unguided agentic optimization is vulnerable to reward hacking, which can inflate
measured speedups without yielding useful kernels~\cite{Hari:2026}. Our manual
inspection is qualitative and targets known shortcut patterns that bypass the
main computation, including PyTorch fallbacks, precomputed or zeroed outputs,
and branches on benchmark values. We do not classify semantics-preserving
changes within a convolution kernel, such as strength reduction for fixed
filter dimensions, as reward hacking. Data-flow invariants additionally reject
shortcuts that violate declared use-site relationships. Neither mechanism is
exhaustive: inspection may miss subtler shortcuts outside known patterns, and
invariants cannot detect behavior outside their vocabulary.

\inlinesection{Compiler and Language Design.} Higher-level DSLs such as Triton
automate layouts and thread scheduling, eliminating several manual
optimizations. Peak performance still requires low-level controls such as inline
assembly, explicit register-class selection, and scheduling
barriers~\cite{AMD:2025a, Triton:2026}. Because compiler stacks often lag behind
new hardware, vendors rely on leaky language abstractions~\cite{Triton:2026}, or
just ship hand-optimized assembly. \system exposes these controls as policies
chosen by LLMs and implemented by the compiler.

\system remains syntactically close to existing DSLs so that LLMs can generalize
from their training data via in-context learning~\cite{Monea:2025}, enabling
adoption of new frontier models without retraining. Better compiler automation
could absorb more low-level decisions, while a more radical LLM-oriented
language could improve token efficiency and support stronger code generation.

\begin{table*}[ht]
\footnotesize\centering
\renewcommand{\arraystretch}{1.2}
\setlength{\tabcolsep}{3pt}
\begin{tabular}{l*{15}{c}}
\toprule
& \multicolumn{5}{c}{GPT-5.6 Sol}
& \multicolumn{5}{c}{DeepSeek V4 Pro}
& \multicolumn{5}{c}{DeepSeek V4 Flash} \\
\cmidrule(lr){2-6}
\cmidrule(lr){7-11}
\cmidrule(lr){12-16}
Method
& Perf & Time & Turns & Tok. & Cost
& Perf & Time & Turns & Tok. & Cost
& Perf & Time & Turns & Tok. & Cost \\
\midrule

\multicolumn{16}{l}{\small\cellcolor{gray!15} GEMM} \\
\system
& 0.89 & 2552/185/107 & 12 & 0.70 & \$6.49
& 0.88 & 2809/191/96 & 17 & 0.93 & \$1.16
& 0.90 & 3947/187/112 & 13 & 1.00 & \$0.61 \\

KernelFalcon
& 0.55 & 1507/81/337 & 12 & 0.14 & \$2.92
& 0.39 & 1327/137/353 & 12 & 0.10 & \$0.23
& 0.49 & 306/21/51 & 8 & 0.05 & \$0.04 \\

KSearch
& 0.36 & 802/127/209 & 12 & 0.16 & \$2.09
& 0.06 & 3399/220/97 & 20 & 0.41 & \$0.81
& 0.30 & 5621/326/117 & 20 & 0.86 & \$0.71 \\

CUDAForge
& 0.07 & 752/294/919 & 11 & 0.11 & \$1.86
& 0.01 & 1789/159/512 & 19 & 0.17 & \$0.34
& 0.03 & 489/238/776 & 17 & 0.13 & \$0.08 \\

Atrex
& 0.51 & 11201/1/465 & 327 & 60.56 & \$46.26
& 0.24 & 9980/0/123 & 524 & 105.54 & \$8.17
& 0.88 & 2033/13/295 & 207 & 22.64 & \$0.56 \\

\addlinespace
\multicolumn{16}{l}{\small\cellcolor{gray!15} Attention} \\
\system
& 0.99 & 1999/16/63 & 6 & 0.44 & \$4.67
& 0.94 & 1175/14/66 & 5 & 0.35 & \$0.51
& 0.94 & 2569/18/67 & 7 & 0.70 & \$0.40 \\

KernelFalcon
& 0.33 & 1570/215/336 & 12 & 0.14 & \$2.93
& 0.09 & 1958/50/175 & 13 & 0.13 & \$0.27
& 0.10 & 523/23/54 & 10 & 0.11 & \$0.07 \\

KSearch
& 0.15 & 1237/121/130 & 19 & 0.31 & \$4.06
& 0.03 & 2891/211/130 & 14 & 0.34 & \$0.68
& 0.03 & 3812/31/133 & 10 & 0.62 & \$0.51 \\

CUDAForge
& 0.001 & 681/271/896 & 11 & 0.12 & \$2.08
& 0.01 & 1648/344/1123 & 14 & 0.20 & \$0.44
& 0.01 & 615/187/314 & 17 & 0.17 & \$0.10 \\

Atrex
& 0.91 & 9477/2/660 & 347 & 49.68 & \$34.55
& 0.68 & 8524/6/153 & 356 & 60.47 & \$4.35
& 1.01 & 4350/6404/1254 & 720 & 112.21 & \$1.57 \\

\addlinespace
\multicolumn{16}{l}{\small\cellcolor{gray!15} Fused MoE} \\
\system
& 0.94 & 2651/17/57 & 9 & 0.75 & \$7.01
& 0.91 & 2400/26/80 & 10 & 0.77 & \$1.10
& 0.92 & 3416/67/224 & 11 & 1.02 & \$0.56 \\

KernelFalcon
& 0.12 & 725/23/54 & 8 & 0.11 & \$1.80
& 0.0002 & 1282/130/321 & 12 & 0.18 & \$0.29
& 0.02 & 764/123/231 & 13 & 0.16 & \$0.10 \\

KSearch
& 0.07 & 2111/113/110 & 14 & 0.39 & \$5.71
& \textbf{\textsc{Fail}} & 5829/484/579 & 20 & 0.83 & \$1.38
& 0.002 & 937/379/367 & 14 & 0.27 & \$0.14 \\

CUDAForge
& 0.02 & 676/272/1375 & 10 & 0.11 & \$2.03
& 0.001 & 1135/156/357 & 19 & 0.16 & \$0.30
& \textbf{\textsc{Fail}} & 4320/461/640 & 20 & 0.70 & \$0.57 \\

Atrex
& 0.32 & 14084/14/670 & 265 & 196.98 & \$146.77
& \textbf{\textsc{Fail}} & 9723/61/2129 & 1170 & 145.34 & \$9.20
& 0.01 & 7706/6/4 & 556 & 193.98 & \$2.71 \\

\bottomrule
\end{tabular}%
\caption{Optimization performance and resource costs across workloads,
frameworks, and models.}
\label{t:token-cost}
\floatrule
\end{table*}

\inlinesection{Generalizing across Hardware Architectures.} Porting \system to a new GPU
requires (\underline{i})~integrating the target toolchain and exposing its
low-level intrinsics and (\underline{ii})~adding hardware-specific optimizations
to the knowledge base. ICRL binds these entries to new kernels without model
retraining, but experts must encode new techniques.

We added NVIDIA Hopper and AMD CDNA4 backends with localized changes. Initial
support for both architectures took approximately one person-week. For CDNA4,
we exposed FP4 intrinsics. For Hopper, we exposed warp-group matrix
multiply-accumulate (WGMMA) and \texttt{mbarrier}. We then added
architecture-specific knowledge entries. NVIDIA Blackwell requires one further
abstraction because its kernels dynamically manage TMEM. We plan to model TMEM
as shared memory using a simple lifetime heuristic: optimized kernels typically
allocate TMEM near entry and
release it near exit. This abstraction would propagate tags through TMEM reads
and writes without modifying the core analysis.

% !Tex Root = ./paper.tex

\section{Evaluation}
\label{s:eval}

Our evaluation answers the following research questions:

\begin{enumerate} [label=\underline{\emph{Q\arabic*.}}]
\item How effectively does \system optimize high-performance GPU kernels?
(\S\ref{sec:exp_q1})
\item How do ICRL and data-flow invariants affect an agent's ability to apply
optimizations?
(\S\ref{sec:exp_q2})
\item How well does \system generalize across kernel families and GPU
architectures? (\S\ref{sec:exp_q3})
\item How do data-flow invariants affect optimization success and cost?
(\S\ref{sec:exp_q4})
\end{enumerate}

\inlinesection{Experimental Environment.} Our primary server has two 64-core
AMD EPYC 9554 CPUs, 2\,TB of DDR5, and eight 192\,GB AMD MI300X GPUs. It runs
Ubuntu 22.04.5, Linux 5.15.0, and ROCm 7.2.2~\cite{AMD:2026a}, whose LLVM 20.0
fork underlies our compiler extensions. Our portability server has two 48-core
Intel Xeon Platinum 8558 CPUs, 2\,TB of DDR5, and eight 141\,GB NVIDIA H200
GPUs. It runs Ubuntu 24.04.1, CUDA 12.8, and driver 570.148.08. We warm up each
kernel for one second and report its mean latency over at least 100 runs within
GPU graphs, pinning each process to the GPU-local CPU socket. We evaluate all agentic frameworks
with GPT-5.6 Sol (GPT)~\cite{OpenAI:2026a}, DeepSeek V4 Pro
(DSv4)~\cite{DeepSeek-AI:2026}, and DeepSeek V4 Flash
(DSv4F)~\cite{DeepSeek-AI:2026}, adapting their prompts for MI300X. For
\system, the selected model drives every Codex-based harness stage, including
planning, lowering, policy evaluation, analysis, and prompt updates.

Turns, Tok., Cost, and Time are averages over attempted problems. Pass@$k$
generation entries are per attempt. A \emph{turn} is an LLM-generation segment
between tool calls. Tok. is in millions, Cost is the API charge in U.S. dollars,
and Time splits LLM/compile/evaluate wall time in seconds. DeepSeek Cost averages
on- and off-peak prices.

Unless noted, throughput is normalized to hipBLASLt for MI300X GEMM, AITER for
MI300X attention and MoE, and the PyTorch reference supplied by KernelBench.
We consider a kernel valid if it compiles, passes all invariant checks, correctness
tests, and a manual reward-hacking review (\S\ref{s:discussion}). 
Only valid kernels contribute to performance. Perf. and Mean are geometric means
of normalized throughput. Min and Max are extrema, and \textgreater1$\times$
counts ratios above parity. Pass@$k$ counts tasks solved within $k$ attempts.
KernelBench generation is a separate first phase that stops at the first
test-passing candidate. Optimization starts from that candidate. Post-hoc
rejection counts as failure without resuming. \textsc{Fail} means no valid
implementation while we report the costs for references. Success requires every
requested transformation.

\begin{figure*}[ht]
\edef\originalcolumnwidth{\the\columnwidth}
\begin{minipage}[t]{\textwidth}
\captionsetup{skip=3pt}
\begin{minipage}[b]{\originalcolumnwidth}
\scriptsize\centering
\renewcommand{\arraystretch}{1.7}
\setlength{\tabcolsep}{2pt}
\begin{tabular*}{\linewidth}{@{\extracolsep{\fill}}lccccccccc@{}}
\toprule
& \rotatebox{90}{\system} & \rotatebox{90}{hipBLASLt} &
\rotatebox{90}{HipKittens} & \rotatebox{90}{AITER} &
\rotatebox{90}{Triton} & \rotatebox{90}{Atrex} &
\rotatebox{90}{KernelFalcon} & \rotatebox{90}{KSearch} &
\rotatebox{90}{CUDAForge} \\
\midrule
\multicolumn{10}{l}{\small\cellcolor{gray!15} Global intrusive changes} \\
\small\hspace{1em}Software pipelining & \checkmark & \checkmark & \checkmark &
\checkmark & \checkmark & \checkmark & \checkmark & - & - \\
\small\hspace{1em}Split K & - & - & - & \checkmark & - & - & - & - & - \\
\small\hspace{1em}MFMA matrix multiplication & \checkmark & \checkmark & \checkmark & \checkmark &
\checkmark & \checkmark & \checkmark & \checkmark & - \\
\small\hspace{1em}Stagger K & \checkmark & \checkmark & \checkmark & \checkmark &
\checkmark & - & - & - & - \\
\small\hspace{1em}Async memcpy & \checkmark & \checkmark & \checkmark & \checkmark &
- & \checkmark & - & - & - \\
\multicolumn{10}{l}{\small\cellcolor{gray!15} Local source changes} \\
\small\hspace{1em}Tile-size specialization & \checkmark & \checkmark & \checkmark &
\checkmark & \checkmark & - & \checkmark & - & - \\
\small\hspace{1em}Bank conflict mitigation & \checkmark & \checkmark & \checkmark &
\checkmark & \checkmark & \checkmark & - & - & \xmark \\
\small\hspace{1em}Vectorized loads & \checkmark & \checkmark & \checkmark &
\checkmark & \checkmark & \checkmark & \checkmark & - & - \\
\small\hspace{1em}Loop unrolling & \checkmark & \checkmark & \checkmark & \checkmark &
\checkmark & \checkmark & \checkmark & \checkmark & - \\
\small\hspace{1em}Workgroup swizzling & \checkmark & \checkmark & \checkmark &
\checkmark & \checkmark & - & \checkmark & - & - \\
\multicolumn{10}{l}{\small\cellcolor{gray!15} ISA-specific optimizations} \\
\small\hspace{1em}HW OOB-guarded loads & \checkmark & \checkmark & \checkmark &
\checkmark & - & \checkmark & - & - & - \\
\small\hspace{1em}Use AGPRs & \checkmark & \checkmark & \checkmark & \checkmark &
- & \checkmark & - & - & - \\
\small\hspace{1em}Explicit instruction scheduling & \checkmark & \checkmark & \checkmark &
\checkmark & - & \checkmark & - & - & - \\
\bottomrule
\end{tabular*}
\captionsetup[table]{position=bottom}
\makeatletter\let\caption@type@warning\@undefined\makeatother
\captionof{table}{GEMM optimizations implemented by \system and other
agentic frameworks using GPT-5.6 Sol, alongside hand-optimized
libraries. \checkmark, -, and \xmark~denote implemented, absent, and incorrectly
implemented optimizations.}
\phantomsection
\label{t:breakdown-gemm}
\end{minipage}\hfill
\begin{minipage}[b]{\originalcolumnwidth}
\centering
\input{figs/build/perf-overall}
\captionof{figure}{Optimization performance on MI300X using GPT-5.6 Sol.
Dashed lines denote parity. Vertical axes are logarithmic.}
\Description{Three grouped bar charts compare normalized throughput for square
BF16 GEMM, GQA flash attention, and fused MoE across five input sizes. Ave is
near the expert reference across all workloads, while agentic baselines are
generally slower.}
\label{f:overall-performance}
\end{minipage}
\par\noindent\floatrule
\end{minipage}
\end{figure*}

\subsection{Kernel Performance}
\label{sec:exp_q1}
\enlargethispage{\baselineskip}

To answer \underline{\emph{Q1}}, we evaluate five
configurations each of three LLM-inference kernels. GEMM multiplies square BF16
matrices with normally distributed entries. Flash attention uses causal BF16
grouped-query attention (GQA) with batch 16, eight query heads, one key-value
head, and head dimension 128,
matching LLaMA-3 70B prefill under eight-way tensor parallelism. Fused MoE uses
DeepSeek-V3.2's block-quantized FP8 configuration~\cite{DeepSeek-AI:2025}:
model dimension 7168, intermediate
dimension 2048, top-$k$ 4, 32 experts per GPU, and expert parallelism of eight.
Each workload starts from a
na\"{\i}ve implementation. Table~\ref{t:token-cost} covers the complete
optimization process, including costs for failed attempts.

Expert-optimized baselines include
hipBLASLt~\cite{AMD:2026}, HipKittens~\cite{Hu:2025}, AITER~\cite{AMD:2025a},
and Triton~\cite{Tillet:2019}. Agentic baselines include Atrex~\cite{Yang:2026},
KernelFalcon~\cite{Wang:2025}, KSearch~\cite{Cao:2026}, and
CUDA\-Forge~\cite{Zhang:2025}. We run \system, KernelFalcon, KSearch, and
CUDA\-Forge for up to ten optimization trials and Atrex for up to five because of
its API cost, reporting the best valid result. Each trial is one optimization
round and may contain multiple LLM turns. These limits may leave some opportunities
unexplored. In our experiments, we adopt hipBLASLt from ROCm 7.2.2, AITER 0.1.10post3,
and the ROCm Triton fork 3.6.0\-+rocm7.2.2.\-git4ed88892. For each workload, the
Triton baseline adopts the faster available Triton implementation from
HipKittens or AITER.
HipKittens GQA lacks MI300X support. We enable available autotuning.

In Figure~\ref{f:overall-performance}, hipBLASLt for GEMM and AITER for
attention and MoE serve as the expert references for each workload. With GPT,
\system's normalized throughput is $0.89$--$0.99\times$ relative to these
references and $1.62$--$1176\times$ relative to uncontaminated agentic
baselines. Atrex uses many more turns than \system and the other agentic
frameworks because it relies on coding agents such as Codex and Claude Code for
long optimization trajectories.

\inlinesection{GEMM.} Table~\ref{t:breakdown-gemm} groups optimizations by
implementation difficulty for an LLM, not performance impact. Global intrusive
changes are hardest because they coordinate control flow, data movement, and
layouts across a kernel. These include pipelining, Split~$K$, MFMA layouts,
Stagger~$K$ for memory-controller hotspots~\cite{AMD:2025b}, and asynchronous
copies. Local changes need only small patches for tile specialization,
bank-conflict mitigation, vectorization, unrolling, or workgroup swizzling.
ISA-specific optimizations require architectural knowledge and low-level
support for hardware OOB protection, AGPRs, and explicit scheduling. AMD
intrinsics cover guarded loads and scheduling, while AGPR allocation requires
assembly or whole-kernel
register management, as in HipKittens.

\system's normalized throughput is $0.89\times$ relative to hipBLASLt. Atrex and
KernelFalcon are the strongest agentic competitors, while among the agentic
baselines, KernelFalcon is the only framework that completes every workload
with every model. Among direct CUDA/HIP frameworks, only
KSearch requests and uses MFMA.

KernelFalcon relies on Triton for pipelining, vectorization, unrolling, and tile
autotuning, but lacks Stagger~$K$ and low-level scheduling. \system's normalized
throughput is $1.62\times$ relative to KernelFalcon and $1.76\times$ relative to
Atrex's CuTe-like FlyDSL~\cite{Li:2026} kernel. Atrex applies most listed
transformations but misses workgroup swizzling, while incomplete FlyDSL
autotuning prevents tile specialization. \system's normalized throughput is
also $1.6\times$ relative to expert Triton.
The evaluated ROCm Triton baseline cannot express asynchronous copies, explicit
AGPR allocation, hardware-guarded loads, or explicit scheduling, which
contribute to this gap. AITER is near parity with hipBLASLt at $M=N=K=1024$, but its
normalized throughput decreases at larger sizes as Split~$K$ becomes less
effective than grid partitioning.

\begin{figure}[!t]
\small\centering
\captionsetup{skip=3pt}
\begin{subfigure}{\linewidth}
\centering
\begin{tikzpicture}[gnuplot]
%% generated with GNUPLOT 6.0p4 (Lua 5.5; terminal rev. Jun 2020, script rev. 120)
\tikzset{every node/.append style={font={\fontsize{7.0pt}{8.4pt}\selectfont}}}
\path (0.000,0.000) rectangle (8.382,3.556);
\gpcolor{color=gp lt color border}
\gpsetlinetype{gp lt border}
\gpsetdashtype{gp dt solid}
\gpsetlinewidth{1.00}
\draw[gp path] (1.424,0.711)--(1.604,0.711);
\node[gp node right] at (1.295,0.711) {$0$};
\draw[gp path] (1.424,1.153)--(1.604,1.153);
\node[gp node right] at (1.295,1.153) {$200$};
\draw[gp path] (1.424,1.595)--(1.604,1.595);
\node[gp node right] at (1.295,1.595) {$400$};
\draw[gp path] (1.424,2.038)--(1.604,2.038);
\node[gp node right] at (1.295,2.038) {$600$};
\draw[gp path] (1.424,2.480)--(1.604,2.480);
\node[gp node right] at (1.295,2.480) {$800$};
\node[gp node center] at (2.271,0.495) {1024};
\node[gp node center] at (3.482,0.495) {2048};
\node[gp node center] at (4.693,0.495) {4096};
\node[gp node center] at (5.903,0.495) {8192};
\node[gp node center] at (7.114,0.495) {16384};
\draw[gp path] (1.424,2.701)--(1.424,0.711)--(7.961,0.711)--(7.961,2.701)--cycle;
\node[gp node right] at (2.567,3.311) {Na\"{\i}ve};
\gpfill{rgb color={0.651,0.808,0.890},color=.!75} (2.696,3.255)--(3.134,3.255)--(3.134,3.367)--(2.696,3.367)--cycle;
\gpcolor{rgb color={0.651,0.808,0.890}}
\draw[gp path] (2.696,3.255)--(3.134,3.255)--(3.134,3.367)--(2.696,3.367)--cycle;
\gpfill{rgb color={0.651,0.808,0.890},color=.!75} (1.770,0.711)--(1.909,0.711)--(1.909,1.018)--(1.770,1.018)--cycle;
\draw[gp path] (1.770,0.711)--(1.770,1.017)--(1.908,1.017)--(1.908,0.711)--cycle;
\gpfill{rgb color={0.651,0.808,0.890},color=.!75} (2.980,0.711)--(3.120,0.711)--(3.120,1.051)--(2.980,1.051)--cycle;
\draw[gp path] (2.980,0.711)--(2.980,1.050)--(3.119,1.050)--(3.119,0.711)--cycle;
\gpfill{rgb color={0.651,0.808,0.890},color=.!75} (4.191,0.711)--(4.330,0.711)--(4.330,1.108)--(4.191,1.108)--cycle;
\draw[gp path] (4.191,0.711)--(4.191,1.107)--(4.329,1.107)--(4.329,0.711)--cycle;
\gpfill{rgb color={0.651,0.808,0.890},color=.!75} (5.402,0.711)--(5.541,0.711)--(5.541,1.161)--(5.402,1.161)--cycle;
\draw[gp path] (5.402,0.711)--(5.402,1.160)--(5.540,1.160)--(5.540,0.711)--cycle;
\gpfill{rgb color={0.651,0.808,0.890},color=.!75} (6.612,0.711)--(6.751,0.711)--(6.751,1.278)--(6.612,1.278)--cycle;
\draw[gp path] (6.612,0.711)--(6.612,1.277)--(6.750,1.277)--(6.750,0.711)--cycle;
\gpcolor{color=gp lt color border}
\node[gp node right] at (2.567,3.086) {$V^{T}$};
\gpfill{rgb color={0.122,0.471,0.706},color=.!75} (2.696,3.030)--(3.134,3.030)--(3.134,3.142)--(2.696,3.142)--cycle;
\gpcolor{rgb color={0.122,0.471,0.706}}
\draw[gp path] (2.696,3.030)--(3.134,3.030)--(3.134,3.142)--(2.696,3.142)--cycle;
\gpfill{rgb color={0.122,0.471,0.706},color=.!75} (1.943,0.711)--(2.082,0.711)--(2.082,1.024)--(1.943,1.024)--cycle;
\draw[gp path] (1.943,0.711)--(1.943,1.023)--(2.081,1.023)--(2.081,0.711)--cycle;
\gpfill{rgb color={0.122,0.471,0.706},color=.!75} (3.153,0.711)--(3.293,0.711)--(3.293,1.058)--(3.153,1.058)--cycle;
\draw[gp path] (3.153,0.711)--(3.153,1.057)--(3.292,1.057)--(3.292,0.711)--cycle;
\gpfill{rgb color={0.122,0.471,0.706},color=.!75} (4.364,0.711)--(4.503,0.711)--(4.503,1.116)--(4.364,1.116)--cycle;
\draw[gp path] (4.364,0.711)--(4.364,1.115)--(4.502,1.115)--(4.502,0.711)--cycle;
\gpfill{rgb color={0.122,0.471,0.706},color=.!75} (5.574,0.711)--(5.714,0.711)--(5.714,1.175)--(5.574,1.175)--cycle;
\draw[gp path] (5.574,0.711)--(5.574,1.174)--(5.713,1.174)--(5.713,0.711)--cycle;
\gpfill{rgb color={0.122,0.471,0.706},color=.!75} (6.785,0.711)--(6.924,0.711)--(6.924,1.285)--(6.785,1.285)--cycle;
\draw[gp path] (6.785,0.711)--(6.785,1.284)--(6.923,1.284)--(6.923,0.711)--cycle;
\gpcolor{color=gp lt color border}
\node[gp node right] at (4.424,3.311) {+Async};
\gpfill{rgb color={0.698,0.875,0.541},color=.!75} (4.553,3.255)--(4.991,3.255)--(4.991,3.367)--(4.553,3.367)--cycle;
\gpcolor{rgb color={0.698,0.875,0.541}}
\draw[gp path] (4.553,3.255)--(4.991,3.255)--(4.991,3.367)--(4.553,3.367)--cycle;
\gpfill{rgb color={0.698,0.875,0.541},color=.!75} (2.116,0.711)--(2.255,0.711)--(2.255,1.139)--(2.116,1.139)--cycle;
\draw[gp path] (2.116,0.711)--(2.116,1.138)--(2.254,1.138)--(2.254,0.711)--cycle;
\gpfill{rgb color={0.698,0.875,0.541},color=.!75} (3.326,0.711)--(3.466,0.711)--(3.466,1.238)--(3.326,1.238)--cycle;
\draw[gp path] (3.326,0.711)--(3.326,1.237)--(3.465,1.237)--(3.465,0.711)--cycle;
\gpfill{rgb color={0.698,0.875,0.541},color=.!75} (4.537,0.711)--(4.676,0.711)--(4.676,1.241)--(4.537,1.241)--cycle;
\draw[gp path] (4.537,0.711)--(4.537,1.240)--(4.675,1.240)--(4.675,0.711)--cycle;
\gpfill{rgb color={0.698,0.875,0.541},color=.!75} (5.747,0.711)--(5.887,0.711)--(5.887,1.326)--(5.747,1.326)--cycle;
\draw[gp path] (5.747,0.711)--(5.747,1.325)--(5.886,1.325)--(5.886,0.711)--cycle;
\gpfill{rgb color={0.698,0.875,0.541},color=.!75} (6.958,0.711)--(7.097,0.711)--(7.097,1.335)--(6.958,1.335)--cycle;
\draw[gp path] (6.958,0.711)--(6.958,1.334)--(7.096,1.334)--(7.096,0.711)--cycle;
\gpcolor{color=gp lt color border}
\node[gp node right] at (4.424,3.086) {+Bank};
\gpfill{rgb color={0.200,0.627,0.173},color=.!75} (4.553,3.030)--(4.991,3.030)--(4.991,3.142)--(4.553,3.142)--cycle;
\gpcolor{rgb color={0.200,0.627,0.173}}
\draw[gp path] (4.553,3.030)--(4.991,3.030)--(4.991,3.142)--(4.553,3.142)--cycle;
\gpfill{rgb color={0.200,0.627,0.173},color=.!75} (2.289,0.711)--(2.428,0.711)--(2.428,1.157)--(2.289,1.157)--cycle;
\draw[gp path] (2.289,0.711)--(2.289,1.156)--(2.427,1.156)--(2.427,0.711)--cycle;
\gpfill{rgb color={0.200,0.627,0.173},color=.!75} (3.499,0.711)--(3.639,0.711)--(3.639,1.260)--(3.499,1.260)--cycle;
\draw[gp path] (3.499,0.711)--(3.499,1.259)--(3.638,1.259)--(3.638,0.711)--cycle;
\gpfill{rgb color={0.200,0.627,0.173},color=.!75} (4.710,0.711)--(4.849,0.711)--(4.849,1.277)--(4.710,1.277)--cycle;
\draw[gp path] (4.710,0.711)--(4.710,1.276)--(4.848,1.276)--(4.848,0.711)--cycle;
\gpfill{rgb color={0.200,0.627,0.173},color=.!75} (5.920,0.711)--(6.060,0.711)--(6.060,1.349)--(5.920,1.349)--cycle;
\draw[gp path] (5.920,0.711)--(5.920,1.348)--(6.059,1.348)--(6.059,0.711)--cycle;
\gpfill{rgb color={0.200,0.627,0.173},color=.!75} (7.131,0.711)--(7.270,0.711)--(7.270,1.355)--(7.131,1.355)--cycle;
\draw[gp path] (7.131,0.711)--(7.131,1.354)--(7.269,1.354)--(7.269,0.711)--cycle;
\gpcolor{color=gp lt color border}
\node[gp node right] at (6.281,3.311) {+Pipe.};
\gpfill{rgb color={0.984,0.604,0.600},color=.!75} (6.410,3.255)--(6.848,3.255)--(6.848,3.367)--(6.410,3.367)--cycle;
\gpcolor{rgb color={0.984,0.604,0.600}}
\draw[gp path] (6.410,3.255)--(6.848,3.255)--(6.848,3.367)--(6.410,3.367)--cycle;
\gpfill{rgb color={0.984,0.604,0.600},color=.!75} (2.462,0.711)--(2.601,0.711)--(2.601,1.489)--(2.462,1.489)--cycle;
\draw[gp path] (2.462,0.711)--(2.462,1.488)--(2.600,1.488)--(2.600,0.711)--cycle;
\gpfill{rgb color={0.984,0.604,0.600},color=.!75} (3.672,0.711)--(3.812,0.711)--(3.812,1.749)--(3.672,1.749)--cycle;
\draw[gp path] (3.672,0.711)--(3.672,1.748)--(3.811,1.748)--(3.811,0.711)--cycle;
\gpfill{rgb color={0.984,0.604,0.600},color=.!75} (4.883,0.711)--(5.022,0.711)--(5.022,1.903)--(4.883,1.903)--cycle;
\draw[gp path] (4.883,0.711)--(4.883,1.902)--(5.021,1.902)--(5.021,0.711)--cycle;
\gpfill{rgb color={0.984,0.604,0.600},color=.!75} (6.093,0.711)--(6.233,0.711)--(6.233,2.090)--(6.093,2.090)--cycle;
\draw[gp path] (6.093,0.711)--(6.093,2.089)--(6.232,2.089)--(6.232,0.711)--cycle;
\gpfill{rgb color={0.984,0.604,0.600},color=.!75} (7.304,0.711)--(7.443,0.711)--(7.443,2.092)--(7.304,2.092)--cycle;
\draw[gp path] (7.304,0.711)--(7.304,2.091)--(7.442,2.091)--(7.442,0.711)--cycle;
\gpcolor{color=gp lt color border}
\node[gp node right] at (6.281,3.086) {+Schedule};
\gpfill{rgb color={0.890,0.102,0.110},color=.!75} (6.410,3.030)--(6.848,3.030)--(6.848,3.142)--(6.410,3.142)--cycle;
\gpcolor{rgb color={0.890,0.102,0.110}}
\draw[gp path] (6.410,3.030)--(6.848,3.030)--(6.848,3.142)--(6.410,3.142)--cycle;
\gpfill{rgb color={0.890,0.102,0.110},color=.!75} (2.635,0.711)--(2.774,0.711)--(2.774,1.672)--(2.635,1.672)--cycle;
\draw[gp path] (2.635,0.711)--(2.635,1.671)--(2.773,1.671)--(2.773,0.711)--cycle;
\gpfill{rgb color={0.890,0.102,0.110},color=.!75} (3.845,0.711)--(3.984,0.711)--(3.984,1.922)--(3.845,1.922)--cycle;
\draw[gp path] (3.845,0.711)--(3.845,1.921)--(3.983,1.921)--(3.983,0.711)--cycle;
\gpfill{rgb color={0.890,0.102,0.110},color=.!75} (5.056,0.711)--(5.195,0.711)--(5.195,2.061)--(5.056,2.061)--cycle;
\draw[gp path] (5.056,0.711)--(5.056,2.060)--(5.194,2.060)--(5.194,0.711)--cycle;
\gpfill{rgb color={0.890,0.102,0.110},color=.!75} (6.266,0.711)--(6.406,0.711)--(6.406,2.255)--(6.266,2.255)--cycle;
\draw[gp path] (6.266,0.711)--(6.266,2.254)--(6.405,2.254)--(6.405,0.711)--cycle;
\gpfill{rgb color={0.890,0.102,0.110},color=.!75} (7.477,0.711)--(7.616,0.711)--(7.616,2.249)--(7.477,2.249)--cycle;
\draw[gp path] (7.477,0.711)--(7.477,2.248)--(7.615,2.248)--(7.615,0.711)--cycle;
\gpcolor{color=gp lt color border}
\draw[gp path] (1.424,2.701)--(1.424,0.711)--(7.961,0.711)--(7.961,2.701)--cycle;
\node[gp node center,rotate=-270.0] at (0.639,1.706) {Effective TFLOPS};
\node[gp node center] at (4.692,0.171) {GQA Flash Attention Optimization \quad Sequence length};
%% coordinates of the plot area
\gpdefrectangularnode{gp plot 1}{\pgfpoint{1.424cm}{0.711cm}}{\pgfpoint{7.961cm}{2.701cm}}
\end{tikzpicture}
%% gnuplot variables
\caption{Optimization performance.}
\label{sf:optimization-performance}
\end{subfigure}
\par\addvspace{3pt}
\begin{subfigure}{\linewidth}
\footnotesize\centering
\renewcommand{\arraystretch}{1.15}
\setlength{\tabcolsep}{1.5pt}
\begin{tabular}{p{0.39\linewidth}p{0.4\linewidth}c}
\toprule
Ablation & Optimization outcome &
TFLOPS \\
\midrule
(\underline{i}) Runtime feedback only & +Async and +Bank ineffective & 190 \\
(\underline{ii}) +KB +invariants, -ICRL &
$\mathbf{V}^{\mathsf T}$, +Async, +Bank, +Schedule attempted (+Pipe not reached) &
255 \\
(\underline{iii}) +KB +ICRL, -invariants &
$\mathbf{V}^{\mathsf T}$ fails & 194 \\
(\underline{iv}) Full \system & All five applied & 588 \\
\bottomrule
\end{tabular}
\caption{Harness ablations.}
\label{sf:optimization-ablation}
\end{subfigure}
\caption{Flash attention optimization. Panel (a) applies stages cumulatively in
legend order. Panel (b) reports geometric-mean effective TFLOPs across sequence lengths.}
\Description{Panel (a) is a grouped bar chart showing effective TFLOPS across
five sequence lengths for six configurations: the na\"{\i}ve kernel followed by
five cumulatively applied flash-attention optimizations. Panel (b) is a table comparing the optimization outcomes and
geometric-mean effective TFLOPS of four harness ablations.}
\label{f:optimization-breakdown}
\floatrule
\end{figure}

\inlinesection{Flash Attention.} \system matches AITER's handwritten assembly.
Atrex reaches $0.91\times$ through the contaminated retrieval\protect{
\footnote{Atrex copies the AITER reference kernel from its knowledge base rather
than deriving it through optimization.}}. Among uncontaminated agentic
baselines, \system's normalized throughput is $6.6\times$ and $1176\times$
relative to KSearch and CUDAForge, respectively, and $3\times$ relative to
KernelFalcon. KSearch and CUDAForge omit matrix cores for $\mathbf{QK}^T$ and
$\mathbf{PV}$. KernelFalcon's Triton kernel neither reorganizes and stages
$\mathbf{V}$ for an on-the-fly transpose nor pipelines and interleaves
instructions. The evaluated ROCm Triton baseline cannot express wavefront
specialization or fine-grained pipeline scheduling. When paired with DSv4F,
Atrex also spends substantial compilation time in a trial-and-error search for a
working compiler configuration.

\inlinesection{Fused MoE.} \system and AITER partition tokens and experts, use
matrix cores, and quantize up-projection intermediates to FP8, enabling FP8
matrix cores with $16\times$ the throughput of scalar units. \system's
normalized throughput is $8\times$ relative to KernelFalcon and $51\times$
relative to CUDAForge. A correct fused MoE kernel must preserve blockwise
quantization, route tokens through sorted expert maps, and group accesses for
coalescing. With DeepSeek V4 Pro, Atrex and KSearch fail to produce stable
kernels.

KernelFalcon uses one stage with DSv4 and two with GPT and DSv4F. Successful
KSearch and CUDAForge kernels also fuse the up projection and activation before
writing a workspace. These two-stage designs omit FP8 requantization before the
down projection and therefore cannot use FP8 matrix cores. They also omit
expert parallelism and full vectorization despite 16-byte alignment.

Table~\ref{t:token-cost} shows that \system remains effective across models,
whereas competing frameworks are more model-sensitive. GPT usually produces
higher-throughput kernels but can generate excessive code, causing
instruction-cache thrashing in CUDAForge attention. DSv4 and DSv4F more often
omit optimizations or choose ineffective strategies. For example, DSv4 gives
KernelFalcon a low-occupancy, single-stage MoE kernel. Attention invariant
checks take only 10--40\,ms, negligible beside LLM inference.

\subsection{Ablation Study of \system's Optimization}
\label{sec:exp_q2}

\begin{table*}[ht]
\footnotesize\centering
\renewcommand{\arraystretch}{1.2}
\setlength{\tabcolsep}{1pt}
\begin{tabular}{l@{\hspace{2pt}}rrrrc@{\hspace{3pt}}rrrrc@{\hspace{3pt}}rrrrrc}
\toprule
& \multicolumn{5}{c}{Syntax reference only}
& \multicolumn{5}{c}{In-context examples}
& \multicolumn{6}{c}{In-context optimization} \\
\cmidrule(lr){2-6}
\cmidrule(lr){7-11}
\cmidrule(lr){12-17}
Dataset--Model
& {Pass@1/3}
& Turns
& Tok.
& Cost
& Time
& {Pass@1/3}
& Turns
& Tok.
& Cost
& Time
& \textgreater1$\times$
& Mean/Min/Max
& Turns
& Tok.
& Cost
& Time \\
\midrule
L1-GPT
& 96/99 & 19.20 & 0.81 & \$0.85 & 156/18/3
& 98/100 & 16.71 & 0.63 & \$0.73 & 158/15/6
& 54 & 1.10/0.12/25.56 & 4.11 & 0.10 & \$0.67 & 319/42/5 \\

L1-DSv4
& 94/96 & 22.45 & 1.21 & \$0.13 & 190/28/14
& 95/95 & 19.69 & 0.76 & \$0.09 & 226/25/35
& 40 & 0.64/0.01/29.23 & 4.18 & 0.14 & \$0.18 & 782/41/18 \\

L1-DSv4F
& 97/100 & 17.71 & 0.94 & \$0.05 & 137/17/4
& 100/100 & 15.64 & 0.70 & \$0.04 & 97/14/6
& 50 & 0.99/0.09/30.45 & 4.52 & 0.25 & \$0.16 & 1481/43/5 \\

\midrule

L2-GPT
& 82/87 & 37.59 & 1.96 & \$1.81 & 289/41/4
& 84/88 & 34.30 & 1.94 & \$1.78 & 312/38/18
& 29 & 1.03/0.07/19.80 & 5.62 & 0.20 & \$1.46 & 605/54/11 \\

L2-DSv4
& 82/86 & 41.60 & 3.25 & \$0.30 & 287/58/23
& 83/86 & 30.78 & 1.82 & \$0.19 & 258/35/61
& 18 & 0.47/0.03/6.52 & 4.92 & 0.17 & \$0.24 & 968/39/14 \\

L2-DSv4F
& 77/86 & 40.07 & 3.79 & \$0.13 & 281/24/14
& 79/84 & 30.31 & 2.63 & \$0.10 & 170/20/13 &
34 & 0.99/0.05/41.66 & 5.59 & 0.34 & \$0.23 & 2043/44/10 \\

\bottomrule
\end{tabular}%
\caption{KernelBench generation under syntax-reference and in-context
prompting, followed by optimization of the in-context results.}
\label{t:kb-gen-opt}
\floatrule
\end{table*}

\begin{table}[ht]
\footnotesize\centering
\renewcommand{\arraystretch}{1.2}
\setlength{\tabcolsep}{4pt}
\begin{tabular}{lrrrrc}
\toprule
Dataset--Model
& Success
& Turns
& Tok.
& Cost
& Time \\
\midrule

\multicolumn{6}{l}{\small\cellcolor{gray!15}No invariants} \\
\hspace{1em}$\widetilde{\text{L1}}$-GPT       & 8  & 7.00 & 0.24 & \$1.74 & 607/45/1 \\
\hspace{1em}$\widetilde{\text{L1}}$-DSv4      & 1  & 4.00 & 0.19 & \$0.28 & 1105/40/2 \\
\hspace{1em}$\widetilde{\text{L1}}$-DSv4F     & 1  & 12.00 & 0.86 & \$0.58 & 4968/99/6 \\
\hspace{1em}$\widetilde{\text{L2}}$-GPT       & 12 & 6.50 & 0.25 & \$1.73 & 719/60/2 \\
\hspace{1em}$\widetilde{\text{L2}}$-DSv4      & 7  & 6.43 & 0.28 & \$0.44 & 1763/71/2 \\
\hspace{1em}$\widetilde{\text{L2}}$-DSv4F     & 7  & 6.57 & 0.37 & \$0.23 & 1963/67/10 \\
\multicolumn{6}{l}{\small\cellcolor{gray!15}With invariants} \\
\hspace{1em}$\widetilde{\text{L1}}$-GPT       & 18 & 6.06 & 0.22 & \$1.54 & 679/40/2 \\
\hspace{1em}$\widetilde{\text{L1}}$-DSv4      & 16 & 5.88 & 0.26 & \$0.43 & 1956/38/2 \\
\hspace{1em}$\widetilde{\text{L1}}$-DSv4F     & 19 & 6.47 & 0.39 & \$0.26 & 2313/43/2 \\
\hspace{1em}$\widetilde{\text{L2}}$-GPT       & 25 & 6.12 & 0.22 & \$1.61 & 656/56/6 \\
\hspace{1em}$\widetilde{\text{L2}}$-DSv4      & 26 & 5.54 & 0.22 & \$0.34 & 1273/43/4 \\
\hspace{1em}$\widetilde{\text{L2}}$-DSv4F     & 23 & 5.74 & 0.37 & \$0.26 & 2100/50/4 \\
\bottomrule
\end{tabular}
\caption{Effect of data-flow invariants on optimizing 60 selected KernelBench
tasks.}
\label{t:kb-opt-invariant}
\floatrule
\end{table}

To answer \underline{\emph{Q2}}, we use GPT and the same five attention
configurations as Q1. We cumulatively apply five optimizations to na\"{\i}ve flash
attention: on-demand $\mathbf{V}^{\mathsf T}$ materialization,
asynchronous loads of $\mathbf{Q}$ and $\mathbf{K}$ (+Async), bank-conflict mitigation
(+Bank), pipelining with wavefront specialization (+Pipe.), and instruction
scheduling (+Schedule). \system also exposes fine-grained register allocation
and data-flow invariants.

Figure~\ref{f:optimization-breakdown}(a) shows throughput increasing
monotonically across all five stages. +Pipe gives the largest incremental gain:
75--116\% across sequence lengths and 100\% by geometric mean. The final
configuration reaches $2.7$--$3.6\times$ Na\"{\i}ve throughput. Effective TFLOPS
use dense-attention FLOPs, including the causally skipped upper triangle.
Figure~\ref{f:optimization-breakdown}(b) isolates harness components while
retaining unit tests and runtime feedback in every configuration. Because each
knowledge base (KB) entry normally pairs an optimization recipe with invariant
templates, the no-invariant ablation uses the same KB after stripping those
templates. In (\underline{i}), feedback prompts +Async and +Bank but cannot
diagnose their ineffective implementations, leaving performance near Na\"{\i}ve.
In (\underline{ii}), the KB provides optimization recipes and invariant
templates, but without ICRL, the planner attempts +Schedule before +Pipe.
Scheduling has no effect, leaving performance at +Bank. In (\underline{iii}),
ICRL orders the steps, but without invariants the agent cannot diagnose and
repair a $\mathbf{V}^{\mathsf T}$ layout that mispairs reduction coordinates, so
performance remains near Na\"{\i}ve. The full system in (\underline{iv})
completes the sequence: $\mathbf{V}^{\mathsf T}$, +Async, +Bank, +Pipe., and
+Schedule.

\subsection{Generality Across Architectures and Kernels}
\label{sec:exp_q3}

To answer \underline{\emph{Q3}}, we evaluate both cross-architecture portability
and generality beyond the production kernels in \S\ref{sec:exp_q1}. For
portability, we add an NVIDIA Hopper backend and use GPT to optimize the same
five configurations of each workload on H200. Against
cuBLASLt~\cite{nvidia_cublaslt_12_8}, vLLM's FlashAttention-3 fork
(b3964b1)~\cite{vllm_flash_attention}, and vLLM's DeepGEMM fork
(a6b593d)~\cite{deepseek_deepgemm}, \system achieves geometric-mean normalized
throughputs of $0.98\times$, $1.00\times$, and $0.99\times$ over the five GEMM,
attention, and fused-MoE configurations, respectively.

To assess cross-kernel generality, we evaluate 100 Level~1 (L1) and 100 Level~2
(L2) KernelBench tasks on MI300X with each model. Because the \system DSL is
absent from their training corpora, Table~\ref{t:kb-gen-opt} compares prompting
with only a compact syntax reference against prompting with in-context examples.
Generation uses only unit-test feedback and stops at the first test-passing
candidate. Among generations using in-context examples, manual review flags
direct PyTorch calls or input-specific shortcuts in 3 L1-DSv4, 6 L2-GPT, 13
L2-DSv4, and 5 L2-DSv4F cases. Optimization
starts from this candidate and runs only for the in-context condition. One
L2-DSv4 optimization ignores instructions and falls back to PyTorch. The
reported pass rates count these outputs as failures. In-context examples improve
Pass@1 while reducing turns, tokens, and API cost.

During optimization, \system applies layout transformations, fusion, vectorized
loads, shared-memory staging, and instruction scheduling. Some kernels remain
below parity because their references call hipBLASLt or MIOpen. DSv4 and DSv4F
have lower aggregate pass rates and mean throughput than GPT, partly because
they more often omit transformations such as im2col~\cite{Chellapilla:2006}.

\subsection{Effectiveness and Cost of Data-Flow Invariants}
\label{sec:exp_q4}

To isolate the impact of data-flow invariants (\underline{\emph{Q4}}), we
compare the full system with an ablation that removes invariant templates and,
consequently, generates no tag functions, assertions, or invariant-violation
feedback. We select KernelBench GEMM and convolution tasks from L1
($\widetilde{\text{L1}}$) and L2 ($\widetilde{\text{L2}}$). Each task requests
MFMA matrix multiplication, software pipelining, and HW OOB-guarded loads.
Success requires applying all three.

Table~\ref{t:kb-opt-invariant} shows that invariants improve success in every
dataset--model pair. They generally reduce turns, tokens, and API cost, while
their effect on wall time is mixed. Thus, invariants consistently make intrusive
transformations more reliable.

% !Tex Root = ./paper.tex

\section{Related Work}
\label{s:related}

\inlinesection{Synthesizing GPU Kernels with LLMs.} Frontier coding
models~\cite{DeepSeek-AI:2026, OpenAI:2026a} combined with agentic
workflows~\cite{Baronio:2026, Cao:2026, Chen:2026, Dai:2026, Dong:2026,
Dong:2026a, Hari:2026, Hu:2026, Ouyang:2025, Wang:2025, Yang:2026, Zhai:2024, Zhang:2025}
have made generating GPU kernels from natural language specifications practical.
These workflows differ along three axes: reward signals, optimization knowledge,
and execution quality. Most use correctness and relative performance gains as
rewards, which are susceptible to reward hacking~\cite{Lange:2025}. The
concurrent $\mu$CUTLASS system instead uses Speed-of-Light guidance, penalizing
claims that exceed hardware capabilities to reduce reward hacking~\cite{Hari:2026}.
Sources include fine-tuning and reinforcement learning~\cite{Su:2025}, world
models~\cite{Cao:2026}, and evolutionary search~\cite{Liao:2026}.

Reliable execution of complex optimizations is essential for meaningful
performance gains on production kernels. \system incorporates data-flow
invariants as dense process rewards that guide optimization. Exposing low-level
hardware and compiler intrinsics further enables optimizations such as
instruction scheduling,
distinguishing \system from systems limited to end-to-end correctness and
performance rewards.

\inlinesection{ICL and ICRL.}
In-context learning (ICL) enables LLMs to adapt to new downstream tasks from
examples provided in the prompt without model parameter
updates~\cite{Monea:2025}. Furthermore, in-context reinforcement learning (ICRL)
extends ICL by treating the prompt as a dynamic learnable policy and updating it
based on reward signals. For example, TextGrad~\cite{Yuksekgonul:2024} provides
a general framework for computing text-based gradients over LLM outputs. In
terms of kernel code generation, KernelBlaster~\cite{Dong:2026a} applies ICRL to
CUDA kernel optimization, maintaining a memory of past optimization trajectories
and using them to improve cross-task generalization. \system builds on these
techniques: we use ICRL to learn planner prompts that bind optimization
knowledge to concrete kernels, with data-flow invariant violations serving as
process rewards alongside runtime performance.

\inlinesection{DSL Designs for GPU Kernels.} GPU DSLs balance productivity
against low-level control through their choice of abstraction. Triton describes
workloads at the block level, hiding explicit memory layouts and synchronization.
Halide~\cite{Ragan-Kelley:2013} and Exo~\cite{Ikarashi:2025} separate
algorithm specification from scheduling policy for tensor workloads.
Tile-based DSLs~\cite{Ding:2025, Hari:2026, NVIDIA:2026, Wang:2026} further
structure programs around tiles as primitives for data movement and matrix
multiplication, shifting implementation complexity to compilers. This
convenience can impose an abstraction tax. Triton extensions such as
Gluon~\cite{Triton:2026} recover performance by exposing direct control over
memory layouts, warp specialization, register allocation, and instruction
scheduling.

The \system DSL combines tile abstractions with explicit compiler controls for
LLM-driven development. Its restricted memory model keeps static analysis
tractable, while compile-time data-flow invariants provide dense structural
feedback during optimization. Its Triton-like syntax lets LLMs reuse knowledge
from training.

\inlinesection{Optimizing Compilers for Tensor Programs.} Tensor
compilers~\cite{Chen:2018, Guan:2025, Li:2023, Ma:2020, Sabne:2020, Wu:2025,
Zhao:2021, Zheng:2020} lower DSLs and tensor programs to GPU binaries.
Their optimizations target high tensor core utilization, efficient memory layout,
software pipelining, operator fusion, and overlapping computation with memory
access. They rely on heuristics and exhaustive search to find efficient
schedules. Note that \system complements these compilers: we expose the hardware SIMT
programming model along with low-level intrinsics in the DSL, and delegate
schedule exploration to LLMs.

\inlinesection{Invariants and Information Flow Systems.} Invariants provide a
concise way to specify and reason about end-to-end properties across
multi-layered systems~\cite{Mai:2013, Tang:2010, Yip:2009}. Information flow
analysis enforces such invariants at the system level.
HiStar~\cite{Zeldovich:2006} assigns security labels to data, defines a partial
order on labels, tracks their propagation, and permits information flow only
along that order. JFlow~\cite{Myers:1999} augments Java with information labels
and enforces information flows at the language level. These prior systems focus
on enforcing security policies. \system adapts the same tracking and enforcement
mechanisms to GPU kernel optimization, using tag propagation to check correct
operand pairings and giving LLMs a cross-layer view of relevant data-flow
constraints.

Formal verification~\cite{Chatterjee:2025, Chen:2016, Klein:2009,
Sigurbjarnarson:2016} can detect invariant violations precisely, but multiplies
development effort by $2$--$6\times$~\cite{Klein:2009} for sequential programs.
Verifying concurrent GPU kernels demands additional modeling
effort~\cite{Betts:2012}. \system instead adopts lightweight
static analysis to report actionable validation failures, enabling the agentic
loop to iterate automatically. The current prototype uses Z3~\cite{Moura:2008} to reason about
layout algebra. A refined version could leverage polyhedral
analysis~\cite{Bondhugula:2008, Grosser:2012, Pradelle:2019} for faster
validation.

% !Tex Root = ./paper.tex

\section{Conclusion}
\label{s:conclusion}

We introduced \system, an agentic framework that uses compile-time data-flow
invariants to make LLM-generated GPU optimizations more reliable. Its tile-based
DSL and static analysis turn invariant violations into concrete witnesses,
guiding agents through intrusive transformations without runtime overhead. With
ICRL, \system reaches near-library performance on AMD MI300X and generalizes to
NVIDIA H200. More broadly, pairing domain abstractions with actionable
correctness feedback offers a path to reliable agentic optimization wherever
correctness depends on cross-layer data flow.

\section*{Acknowledgments}

We thank our shepherd, the anonymous reviewers, and Mark Hill for their
thoughtful feedback. We also thank InnoMatrix for providing the evaluation
servers. This work was partially supported in part by donation grant PD26EG02 to
CSE HKUST, HKUST startup grant R9895, RGC-ECS project 26218024, RGC-NSFC project
CRS\_HKUST\-601/24, and National Natural Science Foundation of China grant
62232015.

\clearpage

\balance
\begingroup
\linespread{1.08}\selectfont
\renewcommand{\bibfont}{\small}
\setlength{\bibsep}{0.8pt}
\bibliographystyle{ave-reference-format}
\bibliography{paper}

@inproceedings{Chellapilla:2006,
	author = {Chellapilla, Kumar and Puri, Sidd and Simard, Patrice},
	series = {IWFHR'06},
	title = {High Performance Convolutional Neural Networks for Document Processing},
	url = {https://inria.hal.science/inria-00112631},
	year = {2006}}

@misc{KernelWiki:2026,
	author = {KernelWiki},
	howpublished = {\url{https://github.com/mit-han-lab/KernelWiki}},
	title = {KernelWiki --- Blackwell \& Hopper Kernel Optimization Knowledge Base},
	year = {2026}}

@misc{DeepSeek-AI:2025,
	archiveprefix = {arXiv},
	author = {DeepSeek-AI},
	eprint = {2512.02556},
	primaryclass = {cs.CL},
	title = {DeepSeek-V3.2: Pushing the Frontier of Open Large Language Models},
	url = {https://arxiv.org/abs/2512.02556},
	year = {2025}}

@inproceedings{Zhang:2025a,
	author = {Shulai Zhang and Ningxin Zheng and Haibin Lin and Ziheng Jiang and Wenlei Bao and Chengquan Jiang and Qi Hou and Weihao Cui and Size Zheng and Li-Wen Chang and Quan Chen and Xin Liu},
	series = {MLSys'25},
	title = {{COMET}: Fine-grained Computation-communication Overlapping for Mixture-of-Experts},
	url = {https://openreview.net/forum?id=fGgQS5VW09},
	year = {2025}}

@inproceedings{Cheng:2026,
	author = {Xinhao Cheng and Zhihao Zhang and Yu Zhou and Jianan Ji and Jinchen Jiang and Zepeng Zhao and Ziruo Xiao and Zihao Ye and Yingyi Huang and Ruihang Lai and Hongyi Jin and Bohan Hou and Mengdi Wu and Yixin Dong and Anthony Yip and Zihao Ye and Songting Wang and Wenqin Yang and Xupeng Miao and Tianqi Chen and Zhihao Jia},
	series = {OSDI'26},
	title = {{MPK}: A Compiler and Runtime for {Mega-Kernelizing} Tensor Programs},
	url = {https://www.usenix.org/conference/osdi26/presentation/cheng},
	year = {2026}}

@misc{Yang:2026,
	archiveprefix = {arXiv},
	author = {Lingyun Yang and Yuxiao Wang and Shenghao Liang and Linfeng Yang and Daocheng Ying and Chunbo You and Rui Zhang and Luping Wang and Yinghao Yu and Guodong Yang and Liping Zhang},
	eprint = {2607.14541},
	primaryclass = {cs.AI},
	title = {Are LLM-Generated GPU Kernels Production-Ready? A Trace-Driven Benchmark and Optimization Agent},
	url = {https://arxiv.org/abs/2607.14541},
	year = {2026}}

@misc{NVIDIA:2026a,
	author = {{NVIDIA}},
	howpublished = {\url{https://docs.nvidia.com/cuda/parallel-thread-execution}},
	title = {Parallel Thread Execution ISA Version 9.3},
	year = {2026}}

@misc{Li:2026,
	author = {Felix Li and Shijie Feng and Carlus Huang and Dewei Wang and Hongxia Yang and Peng Sun and Emad Barsoum},
	howpublished = {\url{https://github.com/ROCm/FlyDSL}},
	title = {FlyDSL},
	year = {2026}}

@misc{DeepSeek-AI:2026,
	archiveprefix = {arXiv},
	author = {DeepSeek-AI and Anyi Xu and Bangcai Lin and Bing Xue and Bingxuan Wang and Bingzheng Xu and Bochao Wu and Bowei Zhang and Chaofan Lin and Chen Dong and Chenchen Ling and Chengda Lu and Chenggang Zhao and Chengqi Deng and Chengyu Hou and Chenhao Xu and Chenze Shao and Chong Ruan and Conner Sun and Damai Dai and Daya Guo and Dejian Yang and Deli Chen and Donghao Li and Dongjie Ji and Erhang Li and Fang Wei and Fangyun Lin and Fangzhou Yuan and Feiyu Xia and Fucong Dai and Guangbo Hao and Guanting Chen and Guoai Cao and Guolai Meng and Guowei Li and Han Yu and Han Zhang and Hanwei Xu and Hao Li and Haofen Liang and Haoling Zhang and Haoming Luo and Haoran Wei and Haotian Yuan and Haowei Zhang and Haowen Luo and Haoyu Chen and Haozhe Ji and Hengqing Zhang and Honghui Ding and Hongxuan Tang and Huanqi Cao and Huazuo Gao and Hui Qu and Hui Zeng and J Yang and JQ Zhu and Jia Luo and Jia Song and Jia Yu and Jialiang Huang and Jialu Cai and Jian Liang and Jiangting Zhou and Jiasheng Ye and Jiashi Li and Jiaxin Xu and Jiewen Hu and Jieyu Yang and Jin Chen and Jin Yan and Jingchang Chen and Jingli Zhou and Jingting Xiang and Jingyang Yuan and Jingyuan Cheng and Jingzi Zhou and Jinhua Zhu and Jiping Yu and Joseph Sun and Jun Ran and Junguang Jiang and Junjie Qiu and Junlong Li and Junmin Zheng and Junxiao Song and Kai Dong and Kaige Gao and Kang Guan and Kexing Zhou and Kezhao Huang and Kuai Yu and Lean Wang and Lecong Zhang and Lei Wang and Leyi Xia and Li Zhang and Liang Zhao and Lihua Guo and Lingxiao Luo and Linwang Ma and Linyan Zhu and Litong Wang and Liyu Cai and Liyue Zhang and Longhao Chen and MS Di and MY Xu and Max Mei and Miaojun Wang and Mingchuan Zhang and Minghua Zhang and Minghui Tang and Mingming Li and Mingxu Zhou and Minmin Han and Ning Wang and Panpan Huang and Panpan Wang and Peixin Cong and Peiyi Wang and Peng Zhang and Qiancheng Wang and Qihao Zhu and Qingyang Li and Qinyu Chen and Qiushi Du and Qiwei Jiang and Rui Tian and Ruifan Xu and Ruijie Lu and Ruiling Xu and Ruiqi Ge and Ruisong Zhang and Ruizhe Pan and Runji Wang and Runqian Chen and Runqiu Yin and Runxin Xu and Ruomeng Shen and Ruoyu Zhang and Ruyi Chen and SH Liu and Shanghao Lu and Shangmian Sun and Shangyan Zhou and Shanhuang Chen and Shaofei Cai and Shaoheng Nie and Shaoqing Wu and Shaoyuan Chen and Shengding Hu and Shengyu Liu and Shiqiang Hu and Shirong Ma and Shiyu Wang and Shuiping Yu and Shunfeng Zhou and Shuting Pan and Shuying Yu and Songyang Zhou and Tao Ni and Tao Yun and Tian Jin and Tian Pei and Tian Ye and Tianle Lin and Tianran Ji and Tianyi Cui and Tianyuan Yue and Tingting Yu and Tun Wang and W Zhang and WL Xiao and Wangding Zeng and Wei An and Weilin Zhao and Wen Liu and Wenfeng Liang and Wenjie Pang and Wenjing Luo and Wenjing Yao and Wenjun Gao and Wenkai Yang and Wenlve Huang and Wenqing Hou and Wentao Zhang and Wenting Ma and Xi Gao and Xiang He and Xiangwen Wang and Xianzu Wang and Xiao Bi and Xiaodong Liu and Xiaohan Wang and Xiaokang Chen and Xiaokang Zhang and Xiaotao Nie and Xiaowen Sun and Xiaoxiang Wang and Xin Cheng and Xin Liu and Xin Xie and Xingchao Liu and Xingchen Liu and Xingkai Yu and Xingyou Li and Xinyu Yang and Xinyu Zhang and Xu Chen and Xuanyu Wang and Xuecheng Su and Xueyin Chen and Xuheng Lin and Xuwei Fu and YC Yan and YQ Wang and YW Ma and Yanfeng Luo and Yang Zhang and Yanhong Xu and Yanru Ma and Yanwen Huang and Yao Li and Yao Li and Yao Xu and Yao Zhao and Yaofeng Sun and Yaohui Wang and Yi Qian and Yi Shao and Yi Yu and Yichao Zhang and Yifan Ding and Yifan Shi and Yijia Wu and Yiliang Xiong and Yiling Ma and Ying He and Ying Tang and Ying Zhou and Yingjia Luo and Yinmin Zhong and Yishi Piao and Yisong Wang and Yixiang Zhang and Yixiao Chen and Yixuan Tan and Yixuan Wei and Yiyang Ma and Yiyuan Liu and Yonglun Yang and Yongqiang Guo and Yongtong Wu and Yu Wu and YuKun Li and Yuan Cheng and Yuan Ou and Yuanfan Xu and Yuanhao Li and Yuduan Wang and Yuehan Yang and Yuer Xu and Yuhan Wu and Yuhao Meng and Yuheng Zou and Yukun Zha and Yunfan Xiong and Yupeng Chen and Yuping Lin and Yuqian Cao and Yuqian Wang and Yushun Zhang and Yuting Yan and Yutong Lin and Yuxian Gu and Yuxiang Luo and Yuxiang You and Yuxuan Liu and Yuxuan Zhou and Yuyang Zhou and Yuzhen Huang and ZF Wu and Zehao Wang and Zehua Zhao and Zehui Ren and Zekai Zhang and Zhangli Sha and Zhe Fu and Zhe Ju and Zhean Xu and Zhenda Xie and Zhengyan Zhang and Zheren Gao and Zhewen Hao and Zhibin Gou and Zhicheng Ma and Zhigang Yan and Zhihong Shao and Zhixian Huang and Zhixuan Chen and Zhiyu Wu and Zhizhou Ren and Zhongyu Wu and Zhuoshu Li and Zhuping Zhang and Zian Xu and Zihao Wang and Zihua Qu and Zihui Gu and Zijia Zhu and Zilin Li and Zipeng Zhang and Ziwei Xie and Ziyi Gao and Ziyi Wan and Zizheng Pan and Zongqing Yao},
	eprint = {2606.19348},
	primaryclass = {cs.CL},
	title = {DeepSeek-V4: Towards Highly Efficient Million-Token Context Intelligence},
	url = {https://arxiv.org/abs/2606.19348},
	year = {2026}}

@misc{OpenAI:2026a,
	author = {OpenAI},
	howpublished = {\url{https://openai.com/index/gpt-5-6}},
	title = {GPT-5.6: Frontier intelligence that scales with your ambition},
	year = {2026}}

@misc{deepseek_deepgemm,
	author = {{DeepSeek-AI}},
	howpublished = {\url{https://github.com/deepseek-ai/DeepGEMM}},
	title = {{DeepGEMM}: A High-Performance Tensor Core Kernel Library},
	year = {2025}}

@misc{nvidia_cublaslt_12_8,
	author = {{NVIDIA}},
	howpublished = {\url{https://docs.nvidia.com/cuda/archive/12.8.0/cublas}},
	title = {{cuBLAS Library Documentation}},
	year = {2025}}

@inproceedings{Zeldovich:2006,
	author = {Nickolai Zeldovich and Silas Boyd-Wickizer and Eddie Kohler and David Mazi{\`e}res},
	series = {OSDI'06},
	title = {Making Information Flow Explicit in {HiStar}},
	url = {https://www.usenix.org/conference/osdi-06/making-information-flow-explicit-histar},
	year = {2006}}

@inproceedings{Ikarashi:2025,
	author = {Ikarashi, Yuka and Qian, Kevin and Droubi, Samir and Reinking, Alex and Bernstein, Gilbert Louis and Ragan-Kelley, Jonathan},
	doi = {10.1145/3669940.3707218},
	isbn = {9798400706981},
	location = {Rotterdam, Netherlands},
	series = {ASPLOS '25},
	title = {Exo 2: Growing a Scheduling Language},
	url = {https://doi.org/10.1145/3669940.3707218},
	year = {2025}}

@inproceedings{Ragan-Kelley:2013,
	author = {Ragan-Kelley, Jonathan and Barnes, Connelly and Adams, Andrew and Paris, Sylvain and Durand, Fr\'{e}do and Amarasinghe, Saman},
	doi = {10.1145/2491956.2462176},
	isbn = {9781450320146},
	location = {Seattle, Washington, USA},
	series = {PLDI '13},
	title = {Halide: a language and compiler for optimizing parallelism, locality, and recomputation in image processing pipelines},
	url = {https://doi.org/10.1145/2491956.2462176},
	year = {2013}}

@misc{Chatterjee:2025,
	archiveprefix = {arXiv},
	author = {Bodhisatwa Chatterjee and Drew Zagieboylo and Sana Damani and Siva Hari and Christos Kozyrakis},
	bibsource = {dblp computer science bibliography, https://dblp.org},
	doi = {10.48550/ARXIV.2511.12294},
	eprint = {2511.12294},
	primaryclass = {cs.SE},
	title = {ProofWright: Towards Agentic Formal Verification of {CUDA}},
	url = {https://doi.org/10.48550/arXiv.2511.12294},
	year = {2025}}

@misc{Dong:2026a,
	archiveprefix = {arXiv},
	author = {Shengjun Kris Dong and Sahil Modi and Dima Nikiforov and Sana Damani and Edward Lin and Siva Kumar Sastry Hari and Christos Kozyrakis},
	bibsource = {dblp computer science bibliography, https://dblp.org},
	doi = {10.48550/ARXIV.2602.14293},
	eprint = {2602.14293},
	primaryclass = {cs.LG},
	title = {Kernel{B}laster: Continual Cross-Task {CUDA} Optimization via Memory-Augmented In-Context Reinforcement Learning},
	url = {https://doi.org/10.48550/arXiv.2602.14293},
	year = {2026}}

@misc{Hari:2026,
	archiveprefix = {arXiv},
	author = {Siva Kumar Sastry Hari and Vignesh Balaji and Sana Damani and Qijing Huang and Christos Kozyrakis},
	eprint = {2603.29010},
	primaryclass = {cs.LG},
	title = {Improving Efficiency of {GPU} Kernel Optimization Agents using a Domain-Specific Language and Speed-of-Light Guidance},
	url = {https://arxiv.org/abs/2603.29010},
	year = {2026}}

@misc{Chen:2026,
	archiveprefix = {arXiv},
	author = {Terry Chen and Zhifan Ye and Bing Xu and Zihao Ye and Timmy Liu and Ali Hassani and Tianqi Chen and Andrew Kerr and Haicheng Wu and Yang Xu and Yu-Jung Chen and Hanfeng Chen and Aditya Kane and Ronny Krashinsky and Ming-Yu Liu and Vinod Grover and Luis Ceze and Roger Bringmann and John Tran and Wei Liu and Fung Xie and Michael Lightstone and Humphrey Shi},
	eprint = {2603.24517},
	primaryclass = {cs.LG},
	title = {{AVO}: Agentic Variation Operators for Autonomous Evolutionary Search},
	url = {https://arxiv.org/abs/2603.24517},
	year = {2026}}

@misc{AMD:2026a,
	author = {AMD},
	howpublished = {\url{https://github.com/ROCm/ROCm}},
	title = {ROCm Software},
	year = {2026}}

@misc{GLM-5-Team:2026,
	archiveprefix = {arXiv},
	author = {GLM-5-Team},
	eprint = {2602.15763},
	primaryclass = {cs.LG},
	title = {{GLM}-5: from Vibe Coding to Agentic Engineering},
	url = {https://arxiv.org/abs/2602.15763},
	year = {2026}}

@misc{AMD:2025b,
	author = {AMD},
	howpublished = {\url{https://rocm.docs.amd.com/en/latest/how-to/rocm-for-ai/inference-optimization/workload.html}},
	title = {{AMD} Instinct {MI300X} workload optimization},
	year = {2025}}

@misc{Hu:2025,
	archiveprefix = {arXiv},
	author = {William Hu and Drew Wadsworth and Sean Siddens and Stanley Winata and Daniel Y. Fu and Ryann Swann and Muhammad Osama and Christopher R{\'e} and Simran Arora},
	eprint = {2511.08083},
	primaryclass = {cs.LG},
	title = {Hip{K}ittens: Fast and Furious {AMD} Kernels},
	url = {https://arxiv.org/abs/2511.08083},
	year = {2025}}

@inproceedings{Shazeer:2017,
	author = {Noam Shazeer and Azalia Mirhoseini and Krzysztof Maziarz and Andy Davis and Quoc Le and Geoffrey Hinton and Jeff Dean},
	series = {ICLR'17},
	title = {Outrageously Large Neural Networks: The Sparsely-Gated Mixture-of-Experts Layer},
	url = {https://openreview.net/forum?id=B1ckMDqlg},
	year = {2017}}

@misc{AMD:2026,
	author = {AMD},
	howpublished = {\url{https://rocm.docs.amd.com/projects/hipBLASLt/en/latest}},
	title = {{hipBLASLt}: General Matrix-Matrix Operations Library for AMD GPUs},
	year = {2026}}

@inproceedings{Cousot:1977,
	author = {Cousot, Patrick and Cousot, Radhia},
	doi = {10.1145/512950.512973},
	isbn = {9781450373500},
	location = {Los Angeles, California},
	series = {POPL'77},
	title = {Abstract interpretation: a unified lattice model for static analysis of programs by construction or approximation of fixpoints},
	url = {https://doi.org/10.1145/512950.512973},
	year = {1977}}

@misc{Lange:2025,
	archiveprefix = {arXiv},
	author = {Robert Tjarko Lange and Qi Sun and Aaditya Prasad and Maxence Faldor and Yujin Tang and David Ha},
	eprint = {2509.14279},
	primaryclass = {cs.SE},
	title = {Towards Robust Agentic {CUDA} Kernel Benchmarking, Verification, and Optimization},
	url = {https://arxiv.org/abs/2509.14279},
	year = {2025}}

@article{Ramalingam:1994,
	address = {New York, NY, USA},
	author = {Ramalingam, G.},
	doi = {10.1145/186025.186041},
	issn = {0164-0925},
	issue_date = {Sept. 1994},
	journal = {ACM Trans. Program. Lang. Syst.},
	month = sep,
	number = {5},
	publisher = {Association for Computing Machinery},
	title = {The undecidability of aliasing},
	url = {https://doi.org/10.1145/186025.186041},
	volume = {16},
	pages = {1467--1471},
	year = {1994}}

@misc{Liao:2026,
	archiveprefix = {arXiv},
	author = {Gang Liao and Hongsen Qin and Ying Wang and Alicia Golden and Michael Kuchnik and Yavuz Yetim and Jia Jiunn Ang and Chunli Fu and Yihan He and Samuel Hsia and Zewei Jiang and Dianshi Li and Uladzimir Pashkevich and Varna Puvvada and Feng Shi and Matt Steiner and Ruichao Xiao and Nathan Yan and Xiayu Yu and Zhou Fang and Roman Levenstein and Kunming Ho and Haishan Zhu and Alec Hammond and Richard Li and Ajit Mathews and Kaustubh Gondkar and Abdul Zainul-Abedin and Ketan Singh and Hongtao Yu and Wenyuan Chi and Barney Huang and Sean Zhang and Noah Weller and Zach Marine and Wyatt Cook and Carole-Jean Wu and Gaoxiang Liu},
	eprint = {2512.23236},
	primaryclass = {cs.LG},
	title = {Kernel{E}volve: Scaling Agentic Kernel Coding for Heterogeneous {AI} Accelerators at {Meta}},
	url = {https://arxiv.org/abs/2512.23236},
	year = {2026}}

@misc{AMD:2025a,
	author = {AMD},
	howpublished = {\url{https://github.com/ROCm/aiter}},
	title = {{AITER}: {AI Tensor Engine for ROCm}},
	year = {2025}}

@inproceedings{Luo:2024,
	author = {Luo, Weile and Fan, Ruibo and Li, Zeyu and Du, Dayou and Wang, Qiang and Chu, Xiaowen},
	doi = {10.1109/IPDPS57955.2024.00064},
	series = {IPDPS'24},
	title = {Benchmarking and Dissecting the Nvidia Hopper {GPU} Architecture},
	url = {https://doi.ieeecomputersociety.org/10.1109/IPDPS57955.2024.00064},
	year = {2024}}

@inproceedings{Huerta:2025,
	author = {Huerta, Rodrigo and Shoushtary, Mojtaba Abaie and Cruz, Jos\'{e}-Lorenzo and Gonzalez, Antonio},
	doi = {10.1145/3725843.3756041},
	isbn = {9798400715730},
	series = {MICRO '25},
	title = {Dissecting and Modeling the Architecture of Modern {GPU} Cores},
	url = {https://doi.org/10.1145/3725843.3756041},
	year = {2025}}

@inproceedings{Lam:1988,
	author = {Lam, Monica},
	doi = {10.1145/53990.54022},
	isbn = {0897912691},
	location = {Atlanta, Georgia, USA},
	series = {PLDI '88},
	title = {Software pipelining: an effective scheduling technique for VLIW machines},
	url = {https://doi.org/10.1145/53990.54022},
	year = {1988}}

@misc{Yuksekgonul:2024,
	archiveprefix = {arXiv},
	author = {Mert Yuksekgonul and Federico Bianchi and Joseph Boen and Sheng Liu and Zhi Huang and Carlos Guestrin and James Zou},
	eprint = {2406.07496},
	primaryclass = {cs.CL},
	title = {Text{G}rad: Automatic "Differentiation" via Text},
	url = {https://arxiv.org/abs/2406.07496},
	year = {2024}}

@inproceedings{Liu:2024,
	author = {Liu, Nelson F. and Lin, Kevin and Hewitt, John and Paranjape, Ashwin and Bevilacqua, Michele and Petroni, Fabio and Liang, Percy},
	doi = {10.1162/tacl_a_00638},
	series = {TACL'24},
	title = {Lost in the Middle: How Language Models Use Long Contexts},
	url = {https://aclanthology.org/2024.tacl-1.9/},
	year = {2024}}

@inproceedings{Monea:2025,
	author = {Giovanni Monea and Antoine Bosselut and Kiant{\'e} Brantley and Yoav Artzi},
	series = {COLM'25},
	title = {{LLM}s Are In-Context Bandit Reinforcement Learners},
	url = {https://openreview.net/forum?id=c0RsezY2D1},
	year = {2025}}

@inproceedings{Quintao-Pereira:2008,
	author = {Quint\~{a}o Pereira, Fernando Magno and Palsberg, Jens},
	doi = {10.1145/1375581.1375609},
	isbn = {9781595938602},
	location = {Tucson, AZ, USA},
	series = {PLDI '08},
	title = {Register allocation by puzzle solving},
	url = {https://doi.org/10.1145/1375581.1375609},
	year = {2008}}

@misc{AMD:2025,
	author = {AMD},
	howpublished = {\url{https://www.amd.com/content/dam/amd/en/documents/instinct-tech-docs/instruction-set-architectures/amd-instinct-mi300-cdna3-instruction-set-architecture.pdf}},
	month = {August},
	title = {{AMD} Instinct {MI300} Instruction Set Architecture},
	year = {2025}}

@inproceedings{Moura:2008,
	author = {de Moura, Leonardo and Bj{\o}rner, Nikolaj},
	series = {TACAS'08},
	isbn = {978-3-540-78800-3},
	title = {{Z3}: An Efficient {SMT} Solver},
	year = {2008}}

@article{Grosser:2012,
	author = {Tobias Grosser and Armin Groesslinger and Christian Lengauer},
	doi = {10.1142/S0129626412500107},
	number = {4},
	journal = {Parallel Processing Letters},
	pages = {1250010},
	title = {{Polly} - Performing polyhedral optimizations on a low-level intermediate representation},
	volume = {22},
	year = {2012}}

@inproceedings{Pradelle:2019,
	author = {Pradelle, Beno{\^\i}t and Meister, Beno{\^\i}t and Baskaran, Muthu and Springer, Jonathan and Lethin, Richard},
	series = {ProTools'18},
	title = {Polyhedral Optimization of {TensorFlow} Computation Graphs},
	year = {2019}}

@inproceedings{Bondhugula:2008,
	author = {Bondhugula, Uday and Hartono, Albert and Ramanujam, J. and Sadayappan, P.},
	doi = {10.1145/1375581.1375595},
	isbn = {9781595938602},
	location = {Tucson, AZ, USA},
	series = {PLDI '08},
	title = {A practical automatic polyhedral parallelizer and locality optimizer},
	url = {https://doi.org/10.1145/1375581.1375595},
	year = {2008}}

@inproceedings{Hawblitzel:2015,
	author = {Hawblitzel, Chris and Howell, Jon and Kapritsos, Manos and Lorch, Jacob R. and Parno, Bryan and Roberts, Michael L. and Setty, Srinath and Zill, Brian},
	doi = {10.1145/2815400.2815428},
	isbn = {9781450338349},
	location = {Monterey, California},
	series = {SOSP '15},
	title = {{I}ron{F}leet: proving practical distributed systems correct},
	url = {https://doi.org/10.1145/2815400.2815428},
	year = {2015}}

@inproceedings{Betts:2012,
	author = {Betts, Adam and Chong, Nathan and Donaldson, Alastair and Qadeer, Shaz and Thomson, Paul},
	doi = {10.1145/2384616.2384625},
	isbn = {9781450315616},
	location = {Tucson, Arizona, USA},
	series = {OOPSLA '12},
	title = {{GPUVerify}: a verifier for {GPU} kernels},
	url = {https://doi.org/10.1145/2384616.2384625},
	year = {2012}}

@inproceedings{Klein:2009,
	author = {Klein, Gerwin and Elphinstone, Kevin and Heiser, Gernot and Andronick, June and Cock, David and Derrin, Philip and Elkaduwe, Dhammika and Engelhardt, Kai and Kolanski, Rafal and Norrish, Michael and Sewell, Thomas and Tuch, Harvey and Winwood, Simon},
	doi = {10.1145/1629575.1629596},
	isbn = {9781605587523},
	location = {Big Sky, Montana, USA},
	series = {SOSP '09},
	title = {{seL4}: formal verification of an {OS} kernel},
	url = {https://doi.org/10.1145/1629575.1629596},
	year = {2009}}

@inproceedings{Chen:2016,
	author = {Haogang Chen and Daniel Ziegler and Tej Chajed and Adam Chlipala and M. Frans Kaashoek and Nickolai Zeldovich},
	series = {USENIX ATC'16},
	title = {Using Crash Hoare Logic for Certifying the {FSCQ} File System},
	url = {https://www.usenix.org/conference/atc16/technical-sessions/presentation/chen_haogang},
	year = {2016}}

@inproceedings{Sigurbjarnarson:2016,
	author = {Helgi Sigurbjarnarson and James Bornholt and Emina Torlak and Xi Wang},
	isbn = {978-1-931971-33-1},
	series = {OSDI'16},
	title = {{Push-Button} Verification of File Systems via Crash Refinement},
	url = {https://www.usenix.org/conference/osdi16/technical-sessions/presentation/sigurbjarnarson},
	year = {2016}}

@inproceedings{Tang:2010,
	author = {Tang, Shuo and Mai, Haohui and King, Samuel T.},
	location = {Vancouver, BC, Canada},
	series = {OSDI'10},
	title = {Trust and protection in the Illinois browser operating system},
	year = {2010}}

@inproceedings{Mai:2013,
	author = {Mai, Haohui and Pek, Edgar and Xue, Hui and King, Samuel Talmadge and Madhusudan, Parthasarathy},
	doi = {10.1145/2451116.2451148},
	isbn = {9781450318709},
	location = {Houston, Texas, USA},
	series = {ASPLOS '13},
	title = {Verifying security invariants in {E}xpress{OS}},
	url = {https://doi.org/10.1145/2451116.2451148},
	year = {2013}}

@misc{OpenAI:2025,
	author = {OpenAI},
	howpublished = {\url{https://github.com/openai/codex}},
	title = {Lightweight coding agent that runs in your terminal},
	year = {2025}}

@inproceedings{Yip:2009,
	author = {Yip, Alexander and Wang, Xi and Zeldovich, Nickolai and Kaashoek, M. Frans},
	doi = {10.1145/1629575.1629604},
	isbn = {9781605587523},
	location = {Big Sky, Montana, USA},
	series = {SOSP '09},
	title = {Improving application security with data flow assertions},
	url = {https://doi.org/10.1145/1629575.1629604},
	year = {2009}}

@inproceedings{Myers:1999,
	author = {Andrew C. Myers},
	month = {January},
	series = {POPL '99},
	title = {J{F}low: practical mostly-static information flow control},
	url = {http://www.cs.cornell.edu/andru/papers/popl99/popl99.pdf},
	year = {1999}}

@inproceedings{Lattner:2021,
	author = {Lattner, Chris and Amini, Mehdi and Bondhugula, Uday and Cohen, Albert and Davis, Andy and Pienaar, Jacques and Riddle, River and Shpeisman, Tatiana and Vasilache, Nicolas and Zinenko, Oleksandr},
	doi = {10.1109/CGO51591.2021.9370308},
	series = {CGO'21},
	title = {{{MLIR}}: Scaling Compiler Infrastructure for Domain Specific Computation},
	year = {2021}}

@inproceedings{Sabne:2020,
	author = {Amit Sabne},
	series = {DEEM'20},
	title = {{XLA}: Compiling Machine Learning for Peak Performance},
	url = {https://research.google/pubs/xla-compiling-machine-learning-for-peak-performance/},
	year = {2020}}

@inproceedings{Ma:2020,
	author = {Lingxiao Ma and Zhiqiang Xie and Zhi Yang and Jilong Xue and Youshan Miao and Wei Cui and Wenxiang Hu and Fan Yang and Lintao Zhang and Lidong Zhou},
	isbn = {978-1-939133-19-9},
	series = {OSDI'20},
	title = {Rammer: Enabling Holistic Deep Learning Compiler Optimizations with {rTasks}},
	url = {https://www.usenix.org/conference/osdi20/presentation/ma},
	year = {2020}}

@inproceedings{Zhao:2021,
	author = {Zhao, Jie and Li, Bojie and Nie, Wang and Geng, Zhen and Zhang, Renwei and Gao, Xiong and Cheng, Bin and Wu, Chen and Cheng, Yun and Li, Zheng and Di, Peng and Zhang, Kun and Jin, Xuefeng},
	doi = {10.1145/3453483.3454106},
	isbn = {9781450383912},
	location = {Virtual, Canada},
	series = {PLDI'21},
	title = {{AKG}: automatic kernel generation for neural processing units using polyhedral transformations},
	url = {https://doi.org/10.1145/3453483.3454106},
	year = {2021}}

@inproceedings{Guan:2025,
	author = {Guan, Yue and Qiang, Xinwei and Pan, Zaifeng and Johnson, Daniels and Fang, Yuanwei and Zhou, Keren and Wang, Yuke and Li, Wanlu and Ding, Yufei and Aziz, Adnan},
	doi = {10.1145/3731569.3764798},
	isbn = {9798400718700},
	location = {Lotte Hotel World, Seoul, Republic of Korea},
	series = {SOSP '25},
	title = {Mercury: Unlocking Multi-{GPU} Operator Optimization for {LLM}s via Remote Memory Scheduling},
	url = {https://doi.org/10.1145/3731569.3764798},
	year = {2025}}

@inproceedings{Wu:2025,
	author = {Wu, Mengdi and Cheng, Xinhao and Liu, Shengyu and Shi, Chunan and Ji, Jianan and Ao, Man Kit and Velliengiri, Praveen and Miao, Xupeng and Padon, Oded and Jia, Zhihao},
	isbn = {978-1-939133-47-2},
	location = {Boston, MA, USA},
	series = {OSDI '25},
	title = {Mirage: a multi-level superoptimizer for tensor programs},
	year = {2025}}

@inproceedings{Li:2023,
	author = {Li, Yijin and Zhao, Jiacheng and Qianqi, Sun and Mai, Haohui and Chen, Lei and Cao, Wanlu and Chen, Yanfan and zhicheng, Li and Liu, Ying and Zhang, Xinyuan and Shi, Xiyu and Zhao, Jie and Xue, Jingling and Cui, Huimin and Feng, XiaoBing},
	series = {MLSys'23},
	title = {{SIRIUS}: Harvesting Whole-Program Optimization Opportunities for {DNN}s},
	url = {https://proceedings.mlsys.org/paper_files/paper/2023/file/3e4e24f7e055320fa54c03f6e816775f-Paper-mlsys2023.pdf},
	year = {2023}}

@inproceedings{Zheng:2020,
	author = {Lianmin Zheng and Chengfan Jia and Minmin Sun and Zhao Wu and Cody Hao Yu and Ameer Haj-Ali and Yida Wang and Jun Yang and Danyang Zhuo and Koushik Sen and Joseph E. Gonzalez and Ion Stoica},
	isbn = {978-1-939133-19-9},
	series = {OSDI'20},
	title = {Ansor: Generating {High-Performance} Tensor Programs for Deep Learning},
	url = {https://www.usenix.org/conference/osdi20/presentation/zheng},
	year = {2020}}

@inproceedings{Chen:2018,
	author = {Chen, Tianqi and Moreau, Thierry and Jiang, Ziheng and Zheng, Lianmin and Yan, Eddie and Shen, Haichen and Cowan, Meghan and Wang, Leyuan and Hu, Yuwei and Ceze, Luis and Guestrin, Carlos and Krishnamurthy, Arvind},
	series = {OSDI'18},
	title = {{TVM}: An Automated End-to-End Optimizing Compiler for Deep Learning},
	url = {https://www.usenix.org},
	year = {2018}}

@misc{Triton:2026,
	author = {Triton},
	howpublished = {\url{https://github.com/triton-lang/triton/tree/main/lib/Dialect/Gluon}},
	title = {Gluon},
	year = {2026}}

@inproceedings{Ding:2025,
	author = {Ding, Yaoyao and Hou, Bohan and Zhang, Xiao and Lin, Allan and Chen, Tianqi and Yu, Cody Hao and Wang, Yida and Pekhimenko, Gennady},
	doi = {10.1145/3760250.3762219},
	isbn = {9798400721656},
	location = {USA},
	series = {ASPLOS '26},
	title = {Tilus: A Tile-Level {GPGPU} Programming Language for Low-Precision Computation},
	url = {https://doi.org/10.1145/3760250.3762219},
	year = {2025}}

@inproceedings{Wang:2026,
	author = {Lei Wang and Yu Cheng and Yining Shi and Zhiwen Mo and Zhengju Tang and Wenhao Xie and Tong Wu and Lingxiao Ma and Yuqing Xia and Jilong Xue and Fan Yang and Zhi Yang},
	series = {ICLR'26},
	title = {TileLang: Bridge Programmability and Performance in Modern Neural Kernels},
	url = {https://openreview.net/forum?id=Jb1WkNSfUB},
	year = {2026}}

@inproceedings{Tillet:2019,
	author = {Tillet, Philippe and Kung, H. T. and Cox, David},
	series = {MAPL'19},
	title = {Triton: an intermediate language and compiler for tiled neural network computations},
	year = {2019}}

@misc{NVIDIA:2026,
	author = {NVIDIA},
	howpublished = {\url{https://github.com/NVIDIA/cutlass}},
	journal = {GitHub repository},
	publisher = {GitHub},
	title = {{CUTLASS}: {CUDA} Templates for Linear Algebra Subroutines and Solvers},
	year = {2026}}

@misc{Wang:2025,
	author = {Laura Wang and PyTorch Team},
	howpublished = {\url{https://pytorch.org/blog/kernelfalcon-autonomous-gpu-kernel-generation-via-deep-agents}},
	title = {Kernel{F}alcon: Autonomous GPU Kernel Generation via Deep Agents},
	year = {2025}}

@inproceedings{Baronio:2026,
	author = {Carlo Baronio and Pietro Marsella and Ben Pan and Simon Guo and Silas Alberti},
	series = {ICLR'26},
	title = {Kevin: Multi-Turn {RL} for Generating {CUDA} Kernels},
	url = {https://openreview.net/forum?id=xu1XwVZtDi},
	year = {2026}}

@inproceedings{Ouyang:2025,
	author = {Anne Ouyang and Simon Guo and Simran Arora and Alex L Zhang and William Hu and Christopher Re and Azalia Mirhoseini},
	series = {ICML'25},
	title = {Kernel{B}ench: Can {LLM}s Write Efficient {GPU} Kernels?},
	url = {https://openreview.net/forum?id=yeoN1iQT1x},
	year = {2025}}

@inproceedings{Zhai:2024,
	author = {Yi Zhai and Sijia Yang and Keyu Pan and Renwei Zhang and Shuo Liu and Chao Liu and Zichun Ye and Jianmin Ji and Jie Zhao and Yu Zhang and Yanyong Zhang},
	isbn = {978-1-939133-40-3},
	series = {OSDI'24},
	title = {Enabling Tensor Language Model to Assist in Generating {High-Performance} Tensor Programs for Deep Learning},
	url = {https://www.usenix.org/conference/osdi24/presentation/zhai},
	year = {2024}}

@inproceedings{Dong:2026,
	author = {Juncheng Dong and Yang Yang and Tao Liu and Yang Wang and Feng Qi and Vahid Tarokh and Kaushik Rangadurai and Shuang Yang},
	series = {ICLR'26},
	title = {{STARK}: Strategic Team of Agents for Refining Kernels},
	url = {https://openreview.net/forum?id=nWaZTH1JMx},
	year = {2026}}

@misc{Hu:2026,
	archiveprefix = {arXiv},
	author = {Bodun Hu and Yoga Sri Varshan V and Saurabh Agarwal and Aditya Akella},
	eprint = {2603.02376},
	primaryclass = {cs.DC},
	title = {{CUCo}: An Agentic Framework for Compute and Communication Co-design},
	url = {https://arxiv.org/abs/2603.02376},
	year = {2026}}

@misc{Zhang:2025,
	archiveprefix = {arXiv},
	author = {Zijian Zhang and Rong Wang and Shiyang Li and Yuebo Luo and Mingyi Hong and Caiwen Ding},
	eprint = {2511.01884},
	primaryclass = {cs.LG},
	title = {Cuda{F}orge: An Agent Framework with Hardware Feedback for {CUDA} Kernel Optimization},
	url = {https://arxiv.org/abs/2511.01884},
	year = {2025}}

@misc{Cao:2026,
	archiveprefix = {arXiv},
	author = {Shiyi Cao and Ziming Mao and Joseph E. Gonzalez and Ion Stoica},
	eprint = {2602.19128},
	primaryclass = {cs.AI},
	title = {{K-Search}: {LLM} Kernel Generation via Co-Evolving Intrinsic World Model},
	url = {https://arxiv.org/abs/2602.19128},
	year = {2026}}

@misc{Su:2025,
	archiveprefix = {arXiv},
	author = {Songqiao Su and Xiaofei Sun and Xiaoya Li and Albert Wang and Jiwei Li and Chris Shum},
	eprint = {2512.02551},
	primaryclass = {cs.LG},
	title = {{CUDA-L2}: Surpassing cuBLAS Performance for Matrix Multiplication through Reinforcement Learning},
	url = {https://arxiv.org/abs/2512.02551},
	year = {2025}}

@misc{Dai:2026,
	archiveprefix = {arXiv},
	author = {Weinan Dai and Hanlin Wu and Qiying Yu and Huan-ang Gao and Jiahao Li and Chengquan Jiang and Weiqiang Lou and Yufan Song and Hongli Yu and Jiaze Chen and Wei-Ying Ma and Ya-Qin Zhang and Jingjing Liu and Mingxuan Wang and Xin Liu and Hao Zhou},
	eprint = {2602.24286},
	primaryclass = {cs.LG},
	title = {{CUDA} Agent: Large-Scale Agentic {RL} for High-Performance {CUDA} Kernel Generation},
	url = {https://arxiv.org/abs/2602.24286},
	year = {2026}}

@misc{dao2023flashattention,
	archiveprefix = {arXiv},
	author = {Dao, Tri},
	eprint = {2307.08691},
	primaryclass = {cs.LG},
	title = {Flashattention-2: Faster attention with better parallelism and work partitioning},
	year = {2023}}

@misc{milakov2018online,
	archiveprefix = {arXiv},
	author = {Milakov, Maxim and Gimelshein, Natalia},
	eprint = {1805.02867},
	primaryclass = {cs.PF},
	title = {Online normalizer calculation for softmax},
	year = {2018}}

@misc{Carlisle:2026,
	archiveprefix = {arXiv},
	author = {Carlisle, Jack and Shah, Jay and Stern, Reuben and VanKoughnett, Paul},
	doi = {10.48550/arXiv.2601.05972},
	eprint = {2601.05972},
	primaryclass = {cs.PL},
	title = {Categorical Foundations for {CuTe} Layouts},
	url = {https://doi.org/10.48550/arXiv.2601.05972},
	year = {2026}}

@misc{vllm_flash_attention,
	author = {{vllm-project}},
	howpublished = {\url{https://github.com/vllm-project/flash-attention}},
	title = {Fast and memory-efficient exact attention},
	year = {2026}}
\endgroup

\end{document}